\documentclass[a4paper,14pt]{article}
\pdfoutput=1 

\usepackage{jheppub} 
\usepackage{siunitx}

\usepackage{bm}

\usepackage[T1]{fontenc} 
\usepackage{csquotes}
\usepackage{cancel} 
\usepackage[normalem]{ulem}
\usepackage{todonotes}
\usepackage{marginnote}
\usepackage{amsthm}
\newtheorem{lemma}{Lemma}

\usepackage{import}
\usepackage{comment}
\usepackage{amsmath}
\usepackage{hyperref}
\usepackage{cleveref}
\usepackage{physics}
\usepackage{siunitx}
\usepackage{graphics}
\usepackage{caption}
\usepackage{subcaption}
\usepackage[compat=1.1.0]{tikz-feynman}
\DeclareMathAlphabet{\pazocal}{OMS}{zplm}{m}{n}
\usepackage{collcell}
\usepackage{colortbl}
\usepackage{listings}
\usepackage{float}
\usepackage{xifthen}
\usepackage{xparse}

\Crefformat{figure}{Fig.\,#2#1#3}
\Crefmultiformat{figure}{Figs.\,#2#1#3}{ and~#2#1#3}{, #2#1#3}{ and~#2#1#3}
\crefrangeformat{figure}{Figs.\,#3#1#4 to~#5#2#6}
\Crefformat{section}{Sec.\,#2#1#3}

\usepackage[acronym]{glossaries}

\usepackage{placeins}

\newacronym{FKS}{FKS}{Frixione-Kunst-Signer}
\newacronym{DGLAP}{DGLAP}{Dokshitzer–Gribov–Lipatov–Altarelli–Parisi}

\newacronym{LHC}{LHC}{Large Hadron Collider}
\newacronym{RHIC}{RHIC}{Relativistic Heavy Ion Collider}

\newacronym{PDF}{PDF}{parton distribution function}
\newacronym{FF}{FF}{fragmentation function}
\newacronym{PS}{PS}{parton shower}
\newacronym{POWHEG}{POWHEG}{Positive Weight Hardest Emission Generator}

\newacronym{IR}{IR}{infrared}
\newacronym{UV}{UV}{ultraviolet}
\newacronym{IRC}{IRC}{infrared collinear}

\newacronym{LO}{LO}{leading-order}
\newacronym{NLO}{NLO}{next-to-leading-order}
\newacronym{NLL}{NLL}{next-to-leading-logarithmic}
\newacronym{NNLO}{NNLO}{next-to-next-to-leading-order}
\newacronym{MiNLO}{MiNLO}{Multi-Scale Next-to-Leading Order}

\newacronym{QGP}{QGP}{quark-gluon plasma}
\newacronym{QCD}{QCD}{quantum chromodynamics}
\newacronym{QED}{QED}{quantum electrodynamics}

\newacronym{AJ}{AJ}{Albright-Jarlskog}

\newcommand{\AJrelation}{AJ relation}

\newcommand{\APFEL}{\texttt{APFEL++}}
\NewDocumentCommand{\F}{O{} O{} m}{\ifthenelse{\isempty{#1}}
	{\ensuremath{F_{#3}^{#2}}}
	{\ensuremath{F_{#3,#1}^{#2}}}}

\usepackage{ulem} 

\usepackage{xcolor}
\usepackage{duckuments}

\title{Heavy-quark contributions to the DIS structure functions \boldmath$F_4$ and \boldmath$F_5$ at NLO in the ACOT scheme
}

\author[a]{E. Spezzano,}
\author[a]{T.~Je\v{z}o,}
\author[a,b]{M.~Klasen,}
\author[c,d]{P.~Risse,}
\author[e]{I.~Schienbein}

\affiliation[a]{Institut für Theoretische Physik, Universität Münster,  Wilhelm-Klemm-Straße 9, 48149 Münster, Germany}
\affiliation[b]{School of Physics, The University of New South Wales, Sydney NSW 2052, Australia}
\affiliation[c]{Department of Physics, Southern Methodist University, Dallas, TX 75275-0175, U.S.A.}
\affiliation[d]{Jefferson Lab, Newport News, VA 23606, U.S.A.}
\affiliation[e]{Laboratoire de Physique Subatomique et de Cosmologie, Université Grenoble-Alpes,CNRS/IN2P3, 53 Avenue des Martyrs, 38026 Grenoble, France}
\emailAdd{edoardo.spezzano@uni-muenster.de}
\emailAdd{tomas.jezo@uni-muenster.de}
\emailAdd{michael.klasen@uni-muenster.de}
\emailAdd{prisse@smu.edu}
\emailAdd{schien@lpsc.in2p3.fr}

\abstract{We compute the contributions of heavy quarks to the deep-inelastic scattering structure functions $F_4$ and $F_5$ at next-to-leading order of perturbative QCD in the ACOT scheme. Both analytic results including the details of the calculation as well as numerical results for the neutral and charged current cases are presented. Our study thus lays the groundwork for future measurements of these two structure functions in experiments such as the SHiP experiment.
}

\preprint{MS-TP-25-14, SMU-PHY-25-05, JLAB-THY-25-4643}

\keywords{Deep-inelastic scattering, structure functions, perturbative QCD, next-to-leading order, heavy quarks, ACOT scheme, SHiP experiment, \texttt{APFEL++} implementation}

\begin{document}
\maketitle

\flushbottom

\section{Introduction}
\label{sec:intro}

Deep-inelastic scattering (DIS) plays a pivotal role in the understanding of the structure of nucleons and nuclei, as it allows to probe these complex objects composed of quarks and gluons with a point-like test particle (a charged lepton or neutrino) at high energy and resolution. Historically, both theoretical and experimental efforts have predominantly focused on the structure functions $F_1$, $F_2$, and $F_3$ \cite{ParticleDataGroup:2024cfk}, while the contributions of the structure functions $F_4$ and $F_5$ have been largely ignored. The primary reason for neglecting the structure functions $F_4$ and $F_5$ lies in the fact that their contributions to the cross section are suppressed by kinematic prefactors proportional to the square of the incoming and outgoing lepton masses. As a result, they are practically relevant only in processes involving heavy leptons, such as the tau.

Recent advances in experimental techniques may open up new avenues for exploring these previously neglected aspects of the nucleon and nuclear structure. In particular, a five-year run of the Search for Hidden Particles (SHiP) experiment at CERN \cite{shipcollaboration2022shipexperimentproposedcern} could provide direct measurements of the structure functions $F_4$ and $F_5$ thanks to its unprecedented sensitivity to interactions of the tau lepton and tau neutrino. Furthermore, seven astrophysical tau neutrino candidate events with visible energies from 20 TeV to 1 PeV have recently been observed with the IceCube Neutrino Observatory \cite{IceCube:2024nhk}, and IceCube/DeepCore are investigating their sensitivity to $F_4$ and $F_5$, which increases at lower neutrino energy due to the kinematic prefactors \cite{MammenAbraham:2022xoc}. These experiments therefore have the potential to significantly enhance our understanding of these structure functions and to provide valuable insights into the non-perturbative dynamics in nucleons and nuclei.

In this paper, we present a comprehensive theoretical analysis of the structure functions $F_4$ and $F_5$ in anticipation of the forthcoming experimental results. Heavy quarks play a crucial role in DIS, since they contribute significantly to the neutral- and charged-current structure functions at low Bjorken-$x$. Their production then also offers valuable insights into the gluon and strange quark distributions in the nucleon, which remain poorly constrained by global fits to inclusive DIS data. Theoretical calculations of heavy-quark contributions present a challenge in perturbative QCD, as they require a consistent treatment of mass effects and the resummation of quasi-collinear logarithms. To address this issue, we adopt the Aivazis-Collins-Olness-Tung (ACOT) scheme \cite{Aivazis:1993kh, Aivazis:1993pi}, which provides a systematic framework for incorporating heavy-quark mass effects within a variable flavor number scheme (VFNS) \cite{Collins:1998rz}. Specifically, we compute the heavy-quark contributions to $F_4$ and $F_5$ up to next-to-leading order (NLO) in the strong coupling, thus providing more precise theoretical predictions to be compared with the upcoming experimental data. It should be noted that the NLO calculation for $F_4$ and $F_5$ has already been performed in the fixed flavor number scheme (FFNS) \cite{jeong2023neutrinocrosssectionsinterface}. At this level of precision, only two processes are theoretically allowed for describing the heavy-quark contribution: quark scattering (QS) and boson-gluon fusion (GF). Together, these processes comprehensively capture the contribution of the heavy quark to the structure functions. Both processes have previously been computed for the structure functions $F_1$, $F_2$, and $F_3$ in the ACOT scheme at NLO \cite{Kretzer:1998ju}. Here, we extend the work in Ref.~\cite{Kretzer:1998ju} to the structure functions $F_4$ and $F_5$. For simplicity, the helicity basis formalism, with which the ACOT scheme was originally formulated, will not be used in this paper. Instead, we will adopt the standard basis, characterized by structure functions $F_1, \dots , F_5$, and thus follow the same approach as Ref.\ \cite{Kretzer:1998ju}. The calculations for both processes were carried out considering a generic ($V-A$) boson interaction to account for both charged and neutral currents using \texttt{FeynCalc} \cite{MERTIG1991345, Shtabovenko_2016,Shtabovenko_2020, Shtabovenko_2025}, a \texttt{Mathematica} package for algebraic calculations in quantum field theory.\\

The remainder of this manuscript is organized as follows:
\begin{itemize}
\item \autoref{sec:kinematics} introduces the kinematic variables relevant for deep-inelastic scattering and presents the hadronic tensor convention adopted in this work together with its decomposition into structure functions.
\item \autoref{sec:quark scattering} focuses on the quark scattering process. We present the calculation of the Born, virtual, and real contributions in $D$ space-time dimensions, followed by the derivation of the semi-inclusive and inclusive results. 
\item \autoref{sec:bgf} focuses on the boson-gluon fusion process. This section covers the definition of the process, its Feynman diagrams, kinematics, and the derivation of both semi-inclusive and inclusive results. 
\item \autoref{sec:ACOT} presents our final analytical results for the inclusive structure functions in the ACOT scheme.
\item \autoref{sec:numerical} presents our numerical results. We compare the individual contributions from the GF and QS channels, illustrate their impact on the $(x, Q^2)$ plane, and discuss the significance of heavy-quark contributions and NLO corrections, in particular in the context of SHiP kinematics.
\item \autoref{sec:conclusions} concludes with a summary of our main findings and a discussion of potential directions for future work.
\end{itemize}

Some technical details of the calculation and the numerical implementation have been relegated to the 
appendices \ref{Appendix: Collinear Frame} -- \ref{appendix: Plus Distribution Treatment}.

\section{DIS with massive leptons}

\label{sec:kinematics}

In order to describe the deep-inelastic scattering processes considered in this work, we adopt the standard kinematic variables and conventions for the hadronic tensor as in Ref.\ \cite{ParticleDataGroup:2024cfk}. The process under study is the scattering of a lepton \(\ell_1(k)\) off a nucleon \(N(P)\), producing a lepton \(\ell_2(k')\) and an inclusive hadronic final state \(X\):
\begin{equation}
\ell_1(k) + N(P) \to \ell_2(k') + X.
\end{equation}
The four-momentum transfer is defined as \(q = k - k'\), and the standard DIS invariants
\begin{equation}
Q^2 = -q^2, \quad x = \frac{Q^2}{2P\cdot q}, \quad y = \frac{P\cdot q}{P\cdot k}
\label{eqn: Bjorken x}
\end{equation}
are introduced, where \(Q^2\) is the virtuality of the exchanged boson, \(x\) is the Bjorken scaling variable, and \(y\) measures the inelasticity of the process. The nucleon mass is denoted by \(M\).

Depending on the exchanged gauge boson ($\gamma,Z,W^\pm$), the leptonic tensor is given by
\begin{align*}
   L_{\mu \nu}^{\gamma} &=2  
  \left( -\frac{1}{2}( (\mu_1-\mu_2)^2+Q^2) g^{\mu\nu} 
  + k^\mu k'^\nu 
  + k'^\mu k^\nu 
  - i \lambda \, \epsilon^{\mu\nu\alpha\beta} k_\alpha k'_\beta
\right),\\
L^{\gamma Z}_{\mu\nu} &= \left(g_V^e + e \lambda g_A^e\right) L^{\gamma}_{\mu\nu}, \\
L^{Z}_{\mu\nu} &= \left(g_V^e + e \lambda g_A^e\right)^2 L^{\gamma}_{\mu\nu}- 4 g_{A}^e  \mu_1 \mu_2 (g_{A}^e+  e \lambda g_{V}^e) g_{\mu \nu}, \\
L^{W}_{\mu\nu} &= \left(1 + e \lambda\right)^2 L^{\gamma}_{\mu\nu}- 4 (1+e \lambda) \mu_1 \mu_2 g_{\mu \nu},
\end{align*}
\noindent
where $g_V^e = -\frac{1}{2} + 2 \sin^2 \theta_W$, \,
$g_A^e = -\frac{1}{2}$, and $\mu_1$, $\mu_2$ are the masses of the incoming and outgoing leptons, respectively. The parameter \(\lambda\) denotes the helicity of the incoming lepton. For incoming charged leptons in unpolarized DIS, the helicities are averaged over; consequently, the antisymmetric term proportional to \(\epsilon^{\mu\nu\alpha\beta}\) cancels. For an incoming {neutrino, which is left-handed in the Standard Model, we set \(\lambda = -1\); for an incoming antineutrino, which is right-handed, we set \(\lambda = +1\).

The hadronic tensor, which encodes the non-perturbative information about the nucleon structure, is conventionally decomposed as
\begin{equation}
W_{j}^{\mu \nu}(P,q) = -g^{\mu \nu} F_{1}^{j} + \frac{P^\mu P^\nu}{P \cdot q} F_{2}^{j} - \frac{i \varepsilon^{\mu \nu \alpha \beta} P_\alpha q_\beta}{2 P \cdot q} F_{3}^{j} + \frac{q^\mu q^\nu}{Q^2} F_{4}^j + \frac{P^\mu q^\nu + q^\mu P^\nu}{2 P \cdot q} F_{5}^j + \frac{P^\mu q^\nu - q^\mu P^\nu}{2 P \cdot q} F_{6}^j,
\end{equation}
where $j= \gamma, \, \gamma Z, \, Z,\, W$ denotes the specific electroweak current contribution (electromagnetic, interference, weak neutral, or charged), and $F_i^{j}(x,Q^2)$ are the structure functions depending on the Lorentz-invariant kinematic variables $x$ and $Q^2$.
The number of structure functions appearing in the hadronic tensor is often reduced in real-life calculations, as $F_6$ is only non-zero if QCD violates time-reversal (or charge-parity) symmetry. 
Whether $F_6$ is actually zero can thus be considered an alternative formulation of the strong CP problem of QCD. The coefficients for $F_4$ and $F_5$, and $F_6$ contain factors of $q_\mu$ and $q_\nu$. This implies that their contractions with the leptonic tensor $L_{\mu\nu}$ are proportional to the lepton masses and thus often neglected \cite{Collins:1984xc}.

The total contraction of the hadronic tensor with the leptonic tensor including the effects of the incoming and outgoing lepton masses $\mu_1$ and $\mu_2$ then leads to the following expressions for the double-differential cross sections:
\begin{align}
\frac{d^2\sigma^{\text{NC}}}{dx\,dy} &= \frac{2\pi y \alpha^2}{Q^4} \sum_{k=\gamma,\,\gamma Z,\,Z} \eta_k \, L^{\mu\nu}_k \, W_{\mu\nu}^{k}, \\
\frac{d^2\sigma^{\text{CC}}}{dx\,dy} &= \frac{2\pi y \alpha^2}{Q^4}  \eta_W \, L^{\mu\nu}_W \, W_{\mu\nu}^{W}.
\end{align}
Here, \(\alpha\) is the fine-structure constant. The overall prefactor \(\eta_k\) takes into account the specific interaction and is given by
\begin{align}
\begin{alignedat}{4}
\eta_\gamma &= 1, \quad\,
& \eta_{\gamma Z} &= \left( \frac{G_F M_Z^2}{2 \sqrt{2} \pi \alpha} \right) \left( \frac{Q^2}{Q^2 + M_Z^2} \right), \\
\eta_Z      &= \eta_{\gamma Z}^2, \quad\,
& \eta_W        &= \frac{1}{2} \left( \frac{G_F M_W^2}{4 \pi \alpha} \right)^2 \left( \frac{Q^2}{Q^2 + M_W^2} \right)^2.
\end{alignedat}
\end{align}
where \(G_F\) is the Fermi constant, and $M_Z$ and \(M_W\) are the masses of the $Z$ and \(W\) bosons.

In the case of unpolarized incoming charged leptons, the kinematic coefficients associated with the structure functions $F_4$ and $F_5$ vanish identically within the $\gamma$ and $\gamma Z$ sectors. Consequently, these structure functions provide no contribution to the unpolarized neutral-current cross section in these channels.\footnote{For polarized incoming leptons, the $\gamma Z$ interference sector becomes accessible, thereby allowing for the measurement of the associated structure functions in principle.} The $ZZ$ exchange channel is therefore the only neutral-current contribution in which $F_4$ and $F_5$ may appear for unpolarized beams. However, in the kinematic regime $Q^2 \ll M_Z^2$, this term is heavily suppressed by the propagator factor $\eta_Z$. Experimentally, the neutral-current cross section is dominated by the pure electromagnetic contribution, with sub-dominant yet non-negligible $\gamma Z$ interference terms dependent solely on $F_{1,2,3}$. As a result, isolating the suppressed $ZZ$ component, and specifically its dependence on $F_{4,5}$, poses a significant challenge in the absence of polarization.

For completeness, assuming equal lepton masses ($\mu_2 = \mu_1$), the $ZZ$ contribution to the differential cross section is given by:
\begin{align}
\frac{d^2\sigma^{ZZ}}{dx\,dy} 
&= \frac{4\pi\alpha^2}{x y Q^2} \, \eta^{ZZ} \Bigg\{ \,
\frac{x y^2 \Big( (g_A^{e \, 2}+g_{V}^{e \, 2}) Q^2 + (6 g_A^{e \,2} - 2 g_{V}^{e \, 2})\mu_1^2 \Big)}{Q^2} \, F_1^{Z} \nonumber \\[6pt]
&\quad - \frac{ \Big( (g_{A}^{e \, 2}+g_{V}^{e \, 2})\big((y-1)Q^2 + M^2 x^2 y^2\big) Q^2 + 4 g_{A}^{e \, 2} M^2 x^2 y^2 \mu_1^2 \Big)}{Q^4} \, F_2^{Z} \nonumber \\[6pt]
&\quad -  g_A^e  g_V^e\, x \, y (y-2) \, F_3^{Z} 
+ \frac{2 g_A^{e\, 2} x y^2 \mu_1^2}{Q^2} \left( F_4^{Z} - \, F_5^{Z} \right)
\Bigg\}. 
\label{eq:ZZ cross section}
\end{align}

Notably, $F_4^Z$ and $F_5^Z$ enter this expression exclusively through the linear combination $(F_4^Z - F_5^Z)$. This implies that even if the $ZZ$ channel were successfully isolated, the individual structure functions could not be disentangled without additional polarization observables. Given the magnitude of the $ZZ$ suppression in realistic kinematic configurations, the extraction of $F_{4,5}$ from unpolarized neutral-current data is practically infeasible. By contrast, the use of polarized lepton beams provides access to the less-suppressed $\gamma Z$ interference channel, offering a more viable pathway to probe these structure functions

For the charged current (\(WW\)) interaction, relevant for (anti)neutrinos and unpolarized charged leptons the cross section reads
\begin{align}
\frac{d^2\sigma^{WW}}{dx\,dy} &= \frac{4\pi\alpha^2}{x y Q^2} \tilde{\eta}\eta^W \Bigg\{ \,
\frac{x y^2 \big(Q^2 + \mu_1^2 + \mu_2^2\big)}{Q^2} \, F_1^{W}  - \frac{2 \Big( (y-1) Q^4 + M^2 x^2 y^2 (Q^2 +  \mu_1^2 + \mu_2^2) \Big)}{Q^4} \, F_2^{W} \nonumber \\[6pt] \quad & \mp\frac{x y \big( (2-y)Q^2 + y (\mu_1^2- \mu_2^2) \big)}{Q^2} \, F_3^{W}  + \frac{x y^2 \Big( \mu_1^4 + (Q^2 - 2 \mu_2^2)\mu_1^2 + \mu_2^2(Q^2 + \mu_2^2) \Big)}{Q^4} \, F_4^{W} \nonumber \\[6pt]
&\quad - \frac{2x y \big( (y-1)\mu_1^2 + \mu_2^2 \big)}{Q^2} \, F_5^{W}
\Bigg\}. 
\label{eqn: CC cross section}
\end{align}
Here, \(\tilde{\eta} = 2\) for (anti)neutrino scattering, and \(\tilde{\eta} = 1\) for unpolarized scattering with charged leptons. The \(\mp\) sign in front of \(F_3^W\) corresponds to neutrino (positive) versus antineutrino (negative) scattering.  
In the above expressions one can see, that terms proportional to $F_4$ and $F_5$ are always accompanied by factors of the lepton masses $\mu_1$ and $\mu_2$, so that in the (lepton-)massless limit only $F_1$, $F_2$, and $F_3$ contribute, recovering the familiar three--structure-function description of DIS.

In order to extract the individual structure functions from the scattering amplitude, it is necessary to introduce appropriate projectors acting on the hadronic tensor. The projectors associated with \(F_4\) and \(F_5\) are defined as
\begin{equation}
\begin{aligned}
P_4^{\mu \nu}(P,q) &= \frac{2Q^2}{\Delta^2} \left( P^2 g^{\mu\nu} + \frac{4}{\Delta^2} \left( \left( \frac{\Delta^2}{2} - 3P^2Q^2 \right) P^\mu P^\nu + 3P^4 q^\mu q^\nu - P^2 (P\cdot q) (P^\mu q^\nu + q^\mu P^\nu) \right) \right), \\
P_5^{\mu \nu}(P,q) &= -\frac{4 (P\cdot q)}{\Delta^2} \Big( - \frac{12}{\Delta^2} \left( Q^2 (P\cdot q) P^\mu P^\nu - P^2 (P\cdot q) q^\mu q^\nu + \left( \frac{\Delta^2}{6} - Q^2 P^2 \right) (P^\mu q^\nu + q^\mu P^\nu) \right) \\
& \hspace{2.1cm} +(P\cdot q) g^{\mu\nu} \Big) ,
\label{eqn:projectors}
\end{aligned}
\end{equation}
where
\begin{equation}
\Delta \equiv \Delta\left( (P+q)^2, P^2, -Q^2 \right), \quad \Delta(a,b,c) = \sqrt{a^2 + b^2 + c^2 - 2ab - 2ac - 2bc}.
\label{eqn:triangle function}
\end{equation}
These projectors satisfy
\begin{equation}
P_i^{\mu\nu}(P,q) W_{\mu\nu}(P,q) = F_i(x,Q^2), \quad i=4,5,
\end{equation}
ensuring the isolation of the desired structure functions. At the parton level, for the quark scattering (QS) and boson-gluon fusion (GF) subprocesses, the corresponding projectors are given by
\begin{equation}
\hat{P}_{i, \rm QS}^{\mu \nu} \equiv  P_i^{\mu \nu}(k_1, q), \quad \hat{P}_{i, \rm GF}^{\mu \nu} \equiv  P_i^{\mu \nu}(p, q),
\label{eq:partonic projectors}
\end{equation}
where \(k_1\) and \(p\) denote the momenta of the incoming quark and gluon, respectively.

It is worth emphasizing that, as shown in \cref{eqn:projectors}, our calculation of the Wilson coefficients employs the standard four-dimensional projectors for the structure functions. In the conventional massless formalism (see, e.g., Ref.~\cite{Furmanski:1981cw}), projectors are formulated in $D$ dimensions to regulate collinear divergences. In our framework, the heavy-quark mass acts as a physical cutoff for such divergences, which validates the use of four-dimensional projectors. At the same time, we maintain a fully $D$-dimensional treatment of the phase space (\textit{cf.} \cref{eqn:phase space}) to regulate soft singularities and to guarantee the correct cancellation of infrared (IR) poles between real and virtual contributions, following the methodology of Ref.~\cite{Kretzer:1998ju}.

\section{Quark scattering}
\label{sec:quark scattering}

To compute the contribution of the quark scattering process
\begin{equation}
B^*(q) + Q_1(k_1) \to Q_2(k_2) [+g(p)]
\end{equation}
to the structure functions, we must account for two key components: real emission of an additional gluon and virtual corrections. Here \( B^* \) represents a virtual boson (\(\gamma, Z, W^\pm\)) with momentum $q$, $k_1$ and $k_2$ represent the quark momenta for quark} with masses $m_1$ and $m_2$, and $g$ is the emitted gluon with momentum $p$. 

\begin{figure}[H]
    \centering
    \includegraphics[scale=0.8]{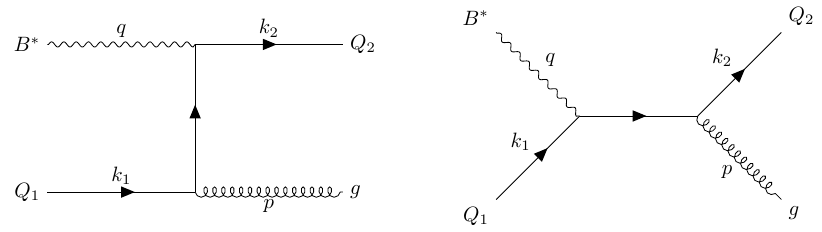}
    \caption{Feynman diagrams relevant for the real contribution at next-to-leading order to the quark scattering process. 
    The emitted gluon is denoted by $g$.}
    \label{fig:Real Emission Diagram}
\end{figure}

\begin{figure}[H]
    \centering
    \includegraphics[scale=0.65]{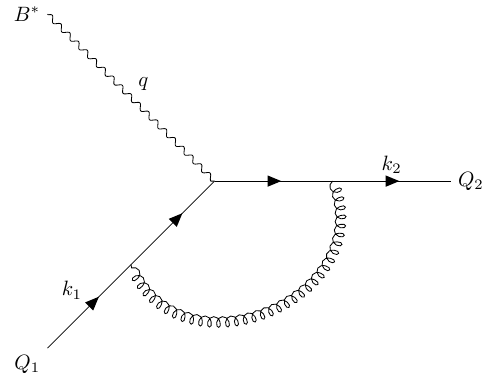}
    \caption{Feynman diagram (triangle vertex correction) relevant for the virtual contribution at next-to-leading order to the quark scattering process.}
    \label{fig:Vertex Correction}
\end{figure}

For a real emission, the relevant Feynman diagrams are shown in \cref{fig:Real Emission Diagram}. These diagrams represent the processes 
involving the presence of an additional gluon in the final state and thus contribute to the scattering process at the next-to-leading order. On the other hand, the virtual contribution is represented by the vertex correction, as illustrated in \cref{fig:Vertex Correction}. In the on-shell renormalization scheme, self-energy insertions on external fermions legs are exactly canceled by the corresponding mass and wave-function renormalization counterterms and therefore do not appear explicitly in the diagram \cite{Peskin:1995ev}. These two contributions must be summed at the cross section level to obtain a consistent result \cite{Kinoshita:1962ur,1964PhRv..133.1549L}.

Since each process involves an external boson, we can write the amplitude in the factorized form 
\begin{equation}
    \mathcal{M} = \varepsilon_{\mu}(q) \mathcal{M}^{\mu},
\end{equation}  
where \( \varepsilon_{\mu}(q) \) denotes the polarization vector of the external gauge boson. Up to phase space integration, the associated partonic tensor is therefore given by 
\begin{equation}
    \hat{W}_{\mu \nu}^{\rm QS} = \mathcal{M}_{\mu} (V, A)\left(\mathcal{M}_{\nu}(V^\prime, A^\prime)\right)^*\, ,
\end{equation}
where $V$ and $A$ denote the usual vector and axial-vector couplings associated with one current, while $V'$ and $A'$ refer to the corresponding couplings of the second current. So, this general expression is applicable to a variety of interactions, such as those involving interference between different gauge bosons (e.g., $\gamma/Z$ exchange)\footnote{We present the amplitude in this general form for completeness, though it is not strictly required for the results of this work.}.

In order to obtain the hadron level structure functions, we use the set of projectors defined in \cref{eq:partonic projectors}.

\subsection{Real contribution}

By applying the projectors specified in \cref{eq:partonic projectors}, the partonic structure functions in the convention of Ref.\ \cite{Kretzer:1998ju} are defined by 
\begin{equation}
\hat{F}_{i}^{\rm QS} \equiv \left(C_F \frac{g_{s}^2}{2}\right)^{-1} \hat{P}_{ i, \, \rm   QS}^{\mu \nu} \overline{\hat{W}_{\mu \nu}^{\rm QS}} , \quad i=4, \, 5,
\end{equation}
where
\begin{equation}
    \overline{\hat{W}_{\mu \nu}^{\rm QS}} = \left( \frac{1}{3} \sum_{\rm color} \right) \left(\frac{1}{2}  \sum_{\rm spins} \right)\hat{W}_{\mu \nu}^{\rm QS}.
    \label{eqn:average tensor boson gluon}
\end{equation}
In this expression, the factor of $1/3$ originates from averaging over the color states of the incoming quark, while the factor of $1/2$ reflects averaging over its spin states. Semi-inclusively, \textit{i.e.}, without integrating over the phase space of the final state quark, we have
\begin{equation}
\begin{aligned}
     \hat{F}_{4}^{\rm QS}(\hat{s}_1,\hat{t}_1) =&  -  16 \frac{ Q^2}{\Delta^{\prime\, 4}} \Bigg\{ q_+ \Big[2 \Delta^2 m_1^2 + \Sigma_{++}^2 \left(7 m_1^2 + 2 m_2^2 - Q^2\right) + \hat{s}_1 \left(-2 \Delta^2 + 14 m_1^2 \Sigma_{++} - \Sigma_{+-}^2\right) + \frac{4 \Delta^2 m_2^2 m_1^2}{\hat{s}_1} \\ &+ \hat{s}_1^2 \left(2 m_1^2 - 3 \Sigma_{+-}\right) + 
    \frac{m_1^2 \hat{s}_1 \left(11 \Delta^2 + 36 m_1^2 m_2^2 - \Sigma_{+-} \Sigma_{++}\right)}{\hat{t}_1} + \frac{4 \Delta^2 m_1^4 \hat{s}_1}{\hat{t}_1^2} +  \frac{4 m_1^4 \hat{s}_1^2 \left(2 \Sigma_{++} - m_1^2\right)}{\hat{t}_1^2} \\ & + \frac{6 m_2^2 m_1^2 \Sigma_{++} \hat{t}_1}{\hat{s}_1}
    + \frac{4 m_1^2 \hat{s}_1^2 \left(m_1^2 + 2 \Sigma_{++}\right)}{\hat{t}_1} + \frac{4 m_1^4 \hat{s}_1^3}{\hat{t}_1^2} + \frac{2 m_1^2 \hat{s}_1^3}{\hat{t}_1} - 3 m_2^2 \Sigma_{+-}^2 + \frac{4 \Delta^2 m_1^2 \Sigma_{++}}{\hat{t}_1}    \\
   &+ \hat{t}_1 \left(\Sigma_{++} \left(4 m_1^2 - \Sigma_{+-}\right) + 4 m_1^2 m_2^2\right) - 2 \hat{s}_1 \hat{t}_1 \Sigma_{+-}  - \hat{s}_1^2 \hat{t}_1 - \hat{s}_1^3 \Big] -2 m_1 m_2 q_- \Big[\frac{\hat{t}_1 \left(\Delta ^2 + 6 m_1^2 m_2^2\right)}{\hat{s}_1} \\& + 6 m_1^2 \left(\frac{m_1^2 \hat{s}_1}{\hat{t}_1} + \hat{s}_1 + \Sigma_{++}\right) + \hat{t}_1 \left(2 \left(m_1^2 + \Sigma_{++}\right) + \hat{s}_1\right)\Big] \Bigg\}
\end{aligned}
\label{eqn:F4 partonic QS}
\end{equation}
and

\begin{equation}
\begin{aligned}
    \hat{F}_{5}^{\rm QS} (\hat{s}_1,\hat{t}_1) =& - 8 \frac{ Q^2}{\hat{x} \Delta^{\prime \, 4}}\Bigg\{q_+ \Big[-2 \Delta^{\prime \, 2} \left(5 m_1^2 + m_2^2 + 2 Q^2\right) + \hat{t}_1 (\hat{s}_1 \left(-7 m_1^2 - m_2^2 - 3 Q^2\right) - 12 m_1^2 m_2^2 + \hat{s}_1^2 \\
    & + \Sigma_{++} (\Sigma_{-+} - 4 \Sigma_{+-})) + \frac{2 \Delta^4 m_2^2}{\hat{s}_1^2} + \hat{s}_1 \left(-3 \Delta^2 - 28 m_1^2 m_2^2 - \Sigma_{-+} \Sigma_{++} - 6 \Sigma_{+-}^2 + 12 \Sigma_{+-} \Sigma_{++}\right) \\& - \Sigma_{+-} \Sigma_{++} \left(7 \Sigma_{++} - 4 m_2^2\right)  + \Delta^2 \left(13 \Sigma_{++} - 2 Q^2\right) +  \frac{2 \Delta^4 \Sigma_{++}}{\hat{t}_1 \hat{s}_1}+ \frac{ 2 m_1^2 \hat{s}_1^3 \left(\Sigma_{+-} \hspace{-0.1em} -  \hspace{-0.1em} 2 m_1^2\right)}{\hat{t}_1^2}
    \\& +  \frac{4 \Delta^2 \left(2 \Delta^2 - m_1^2 \Sigma_{++} + 6 m_1^2 m_2^2\right) + \hat{s}_1^2 \Big(-2 \Delta^2 - 20 m_1^2 m_2^2 + 4 \Sigma_{+-}^2 + 9 \Sigma_{+-} \Sigma_{++}\Big) }{\hat{t}_1} \\
    &+ \frac{\hat{s}_1 \left(\Sigma_{++} \left(9 \Delta^2 - 2 m_1^2 \Sigma_{+-} + 4 \Sigma_{+-}^2\right) - 6 \Delta^2 m_1^2\right) + \hat{s}_1^3 \left(5 \Sigma_{+-} - 4 m_1^2\right) + \hat{s}_1^4}{\hat{t}_1} \\
    &+ \frac{\Delta^2 \left(5 \Delta^2 + 8 m_1^2 m_2^2 - 3 \Sigma_{-+} \Sigma_{++}\right) + \hat{t}_1 \left(\Delta^2 \Sigma_{++} + 2 m_2^2 \left(\Delta^2 - 3 \Sigma_{+-} \Sigma_{++}\right)\right) }{\hat{s}_1} \\
    &+ \frac{2 \Delta^4 m_1^2 + 2 m_1^2 \hat{s}_1^2 \left(\Delta^2 + 2 \Sigma_{++} \left(\Sigma_{+-} - m_1^2\right) + 4 m_1^4\right) + 2 \Delta^2 m_1^2 \hat{s}_1 \left(3 \Sigma_{++} \hspace{-0.1em} -  \hspace{-0.1em} 2 m_1^2\right) }{\hat{t}_1^2}  \Big] \\
    & +2 m_1 m_2 q_{-} \Big[2 \left(3 \Sigma_{+-} \Sigma_{++} - \Delta^2\right) + \frac{6 m_1^2 \hat{s}_1 \left(\hat{s}_1 + \Sigma_{+-}\right)}{\hat{t}_1} + 4 \hat{s}_1 \left(-m_1^2 + \hat{s}_1 + 2 \Sigma_{++} + \hat{t}_1\right)  \\ & +    2 \hat{t}_1 \left(3 m_2^2 + \Sigma_{+-}\right) 
   + \frac{\hat{t}_1 \left(\Delta^2 - 3 \Sigma_{-+} \Sigma_{++}\right)}{\hat{s}_1}\Big] \Bigg\},
\end{aligned}
\label{eqn:F5 partonic QS}
\end{equation}
where we have defined the quantities
\begin{equation}
  q_{\pm} = V V^\prime \pm A A^\prime , \quad  \Sigma_{\pm \pm} = Q^2 \pm m_2^2 \pm m_1^2, \quad  \hat{t}_1=(k_2 - q)^2 -m_{1}^2  , \quad \hat{s}_1= (k_1 + q)^2-m_{2}^2
  \label{eqn:s1 expression}
\end{equation}
and
\begin{equation}
  \Delta \equiv \Delta(m_1^2, m_2^2, -Q^2), \quad \Delta^{\prime} \equiv \Delta( m_1^2,\hat{s}_1+m_2^2, -Q^2),
\end{equation}
with the $\Delta$-function defined in \cref{eqn:triangle function}. Finally, the variable $\hat{x}$ appearing in $\hat{F}_{5}^{\rm QS}$ can be expressed in terms of $\hat{s}_1$ using the relation
\begin{equation}
    \hat{s}_1 = m_1^2-m_{2}^2+Q^2 \left( \frac{1}{\hat{x}} - 1 \right).
\end{equation}
As a check, we have also recalculated the corresponding expressions for the structure functions $\hat F_{1,2,3}^{\rm QS}(\hat{s}_1,\hat{t}_1)$ 
and find full agreement with the results in appendix A of Ref.~\cite{Kretzer:1998nt}.

To obtain the inclusive structure functions, we integrate the semi-inclusive ones over the polar angle $\theta_{\rm CM}$ between $\vec{k}_2$ and $\vec{k}_1$ in the center-of-mass frame with
\begin{equation}
    \hat{F}_{i}^{\rm QS} ( \hat{s_1}) \equiv \int_{0}^{1} \hat{F}_{i}^{\rm QS} \left( \hat{s}_1, \hat{t}_1 (y)\right) \, dy,
\end{equation}
where the integration variable $y$ is defined as
\begin{equation}
     y = \frac{1}{2} \left( 1+ \cos \theta_{\rm CM} \right) 
\end{equation}
and $\hat{t}_1(y)$ is given by
\begin{equation}
     \hat{t}_{1}(y)=\frac{\hat{s}_1}{\hat{s}_1+m_2^2} \Delta^{\prime} (y-y_0)
\end{equation}
with $ y_0=[ 1+ (\Sigma_{++}+ \hat{s}_1)/ \Delta^{\prime}]/2$. Explicitly, the inclusive results can be derived from the semi-inclusive expressions by the following substitutions in \cref{eqn:F4 partonic QS,eqn:F5 partonic QS}:
\begin{align}
    \frac{1}{\hat{t}_1^2} \rightarrow \frac{\hat{s}_1+m_2^2}{m_1^2 \hat{s}_1^2}, \quad \frac{1}{\hat{t}_1}  \rightarrow \frac{\hat{s}_1+m_2^2}{\hat{s}_1 \Delta^{\prime}} \mathcal{L} , \quad \hat{t}_1 \rightarrow  -\frac{\hat{s}_1 (\Sigma_{++}+\hat{s}_1)}{2(\hat{s}_1+m_{2}^2)}
    \label{eqn:from semi-inclusive to inclusive}.
\end{align}
Here, $\mathcal{L}$ is defined as
\begin{equation}
     \mathcal{L} \equiv \ln \left( \frac{\Sigma_{++}+ \hat{s}_1- \Delta^{\prime}}{{\Sigma_{++}+ \hat{s}_1+\Delta^{\prime}}} \right).
     \label{eqn:log term}
\end{equation}

\subsection{Virtual contribution}

Calculating the vertex correction in \cref{fig:Vertex Correction} in $D=4-2 \epsilon$ dimensions and using the Passarino-Veltman decomposition \cite{Passa_Velt}, we obtain 
\begin{equation}
\begin{aligned}
 \Gamma_0^{\mu} & =C_F \frac{\alpha_s}{4 \pi} \frac{1}{\Gamma(1-\epsilon)}\left(\frac{Q^2}{4 \pi \mu^2}\right)^{-\epsilon}\left(C_{0,-} \gamma^\mu L_5+C_{+} \gamma^\mu R_5\right. \\
& \left.+C_{1,-} m_2 k_1^\mu L_5+C_{1,+} m_1 k_1^\mu R_5+C_{q,-} m_2 q^\mu L_5+C_{q,+} m_1 q^\mu R_5\right)
\label{eqn:vertex correction analitical expression}
\end{aligned}
\end{equation}
with $ L_5=(V- A \gamma_5)$ and $R_5= (V+A \gamma_5)$. The coefficients read
\begin{equation}
\begin{aligned}
&C_{0,-}=\frac{1}{\epsilon}\left(1+\Sigma_{++} I_1\right)+\left[\frac{\Delta^2}{2 Q^2}+\Sigma_{++}\left(1+\ln \left(\frac{Q^2}{\Delta}\right)\right)\right] I_1\\
& +\frac{1}{2} \ln \left(\frac{Q^2}{m_1^2}\right)+\frac{1}{2} \ln \left(\frac{Q^2}{m_2^2}\right)+\frac{m_2^2-m_1^2}{2 Q^2} \ln \left(\frac{m_1^2}{m_2^2}\right)+\frac{\Sigma_{++}}{\Delta} \\
& \times\left\{\frac{1}{2} \ln ^2\left|\frac{\Delta-\Sigma_{+-}}{2 Q^2}\right|+\frac{1}{2} \ln ^2\left|\frac{\Delta-\Sigma_{-+}}{2 Q^2}\right|-\frac{1}{2} \ln ^2\left|\frac{\Delta+\Sigma_{+-}}{2 Q^2}\right|-\frac{1}{2} \ln ^2\left|\frac{\Delta+\Sigma_{-+}}{2 Q^2}\right|\right. \\
& \left.-\operatorname{Li}_2\left(\frac{\Delta-\Sigma_{+-}}{2 \Delta}\right)-\operatorname{Li}_2\left(\frac{\Delta-\Sigma_{-+}}{2 \Delta}\right)+\mathrm{Li}_2\left(\frac{\Delta+\Sigma_{+-}}{2 \Delta}\right)+\mathrm{Li}_2\left(\frac{\Delta+\Sigma_{-+}}{2 \Delta}\right)\right\},\\
&C_{+}  =2 m_1 m_2 I_1, \\
&C_{1,-}  =-\frac{1}{Q^2}\left[\Sigma_{+-} I_1+\ln \left(\frac{m_1^2}{m_2^2}\right)\right], \\
&C_{1,+}  =-\frac{1}{Q^2}\left[\Sigma_{-+} I_1-\ln \left(\frac{m_1^2}{m_2^2}\right)\right], \\
&C_{q,-}  =\frac{1}{Q^4}\left[\left(\Delta^2-2 m_1^2 Q^2\right) I_1-2 Q^2+\Sigma_{+-} \ln \left(\frac{m_1^2}{m_2^2}\right)\right], \\
& C_{q,+}  =\frac{1}{Q^4}\left[\left(-\Delta^2+2 m_2^2 Q^2-\Sigma_{-+} Q^2\right) I_1+2 Q^2+\left(\Sigma_{-+}+Q^2\right) \ln \left(\frac{m_1^2}{m_2^2}\right)\right]
\end{aligned}
\label{eqn:VirtualCoefficients}
\end{equation}
with
\begin{equation}
I_1=\frac{1}{\Delta} \ln \left(\frac{\Sigma_{++}+\Delta}{\Sigma_{++}-\Delta}\right).
\end{equation}
The renormalized vertex is given by
\begin{equation}
    \Gamma_{\rm R}^{\mu}= \gamma^{\mu} L_5 (Z_1-1)+ \Gamma_{0}^{\mu},
\end{equation}
where
\begin{equation}
Z_1=1+C_F \frac{\alpha_s}{4 \pi} \frac{1}{\Gamma(1-\epsilon)}\left(\frac{Q^2}{4 \pi \mu^2}\right)^{-\epsilon}\left[-\frac{3}{\epsilon}-\frac{3}{2} \ln \left(\frac{Q^2}{m_1^2}\right)-\frac{3}{2} \ln \left(\frac{Q^2}{m_2^2}\right)-4\right].
\end{equation}
The only coefficient that changes after renormalization is $C_{0,-}$, which is replaced by
\begin{equation}
\begin{aligned}
C_{R,-} & =-\frac{1}{\epsilon}\left(2-\Sigma_{++} I_1\right)+\left[\frac{\Delta^2}{2 Q^2}+\Sigma_{++}\left(1+\ln \left(\frac{Q^2}{\Delta}\right)\right)\right] I_1 \\
& +\frac{m_2^2-m_1^2}{2 Q^2} \ln \left(\frac{m_1^2}{m_2^2}\right)-\ln \left(\frac{Q^2}{m_1^2}\right)-\ln \left(\frac{Q^2}{m_2^2}\right)-4+\frac{\Sigma_{++}}{\Delta} \\
& \times\left\{\frac{1}{2} \ln ^2\left|\frac{\Delta-\Sigma_{+-}}{2 Q^2}\right|+\frac{1}{2} \ln ^2\left|\frac{\Delta-\Sigma_{-+}}{2 Q^2}\right|-\frac{1}{2} \ln ^2\left|\frac{\Delta+\Sigma_{+-}}{2 Q^2}\right|-\frac{1}{2} \ln ^2\left|\frac{\Delta+\Sigma_{-+}}{2 Q^2}\right|\right. \\
& \left.-\operatorname{Li}_2\left(\frac{\Delta-\Sigma_{+-}}{2 \Delta}\right)-\operatorname{Li}_2\left(\frac{\Delta-\Sigma_{-+}}{2 \Delta}\right)+\operatorname{Li}_2\left(\frac{\Delta+\Sigma_{+-}}{2 \Delta}\right)+\operatorname{Li}_2\left(\frac{\Delta+\Sigma_{-+}}{2 \Delta}\right)\right\}.
\end{aligned}
\label{eqn:CRminus}
\end{equation}
These results are in full agreement with Ref. \cite{Kretzer:1998ju}.

\subsection{Combined contributions}

To obtain the final expression for the partonic structure functions, we start by integrating the real semi-inclusive contribution over the phase space, followed by the addition of the virtual contribution. In \(D=4 - 2\epsilon\) dimensions, the phase space is given by  \cite{PhysRevD.23.56}
\begin{equation}
d\Pi = \frac{1}{8\pi} \left( \frac{1}{\hat{s}_1^{1 + 2\epsilon}} \right) \frac{\hat{s}_1^2}{\hat{s}_1 + m_2^2} \frac{1}{\Gamma(1 - \epsilon)} \left( \frac{\mu^2}{4 \pi (\hat{s}_1 + m_2^2)} \right)^{-\epsilon} y^{-\epsilon} (1 - y)^{-\epsilon} \, dy,
\label{eqn:phase space}
\end{equation}
where we included the renormalization scale $\mu^2$. The term inside the first parentheses can be reformulated using the identity \cite{Altarelli:1979ub}
\begin{equation}
\frac{1}{\hat{s}_1^{1 + 2\epsilon}} = - \frac{1}{\Delta^{1+2\epsilon}}\frac{1}{2 \epsilon} \delta(1- \xi^\prime)+\frac{1}{\hat{s}_1} \frac{1-\xi^\prime}{(1-\xi^\prime)_+}+ \mathcal{O}(\epsilon),
\label{eqn:identity}
\end{equation}
where \(\xi' \equiv \chi/\xi\) with 
\begin{equation}
\chi = \frac{x}{2Q^2} (\Sigma_{+-} + \Delta),
\label{eqn: Chi definition}
\end{equation}
with $x$ being the Bjorken-$x$ variable defined in \cref{eqn: Bjorken x}. 
The identity in \cref{eqn:identity} follows from the expression for \(\hat{s}_1\), which can be written as
\begin{equation}
\hat{s}_1(\xi') = \frac{1 - \xi'}{2\xi'} \left[ (\Delta - \Sigma_{+-}) \xi' + \Delta + \Sigma_{+-} \right],
\end{equation}
derived by combining the expression for \(\hat{s}_1\) given in \cref{eqn:s1 expression} and the scalar products in \cref{scalar products}. 
Thus we obtain the expression
\begin{equation}
\label{eqn:FhatQSatNLOnoMixing}
\hat{\mathcal{F}}_i^{\rm QS, \, NLO}(\xi^\prime) \equiv  \pi \int_0^1 d\Pi \, \hat{F}_{i}^{\rm QS}(\hat{s}_1, \hat{t}_1) + N_i V_i \delta(1 - \xi') =   N_i (S_i + V_i) \delta(1 - \xi') + \frac{1}{8} \frac{1 - \xi'}{(1 - \xi')_+} \frac{\hat{s}_1}{\hat{s}_1 + m_2^2} \hat{F}_{i}^{\rm QS} (\hat{s}_1),
\end{equation}
where \(\hat{F}_i(\hat{s}_1)\) are the inclusive structure functions obtained using the substitutions reported in \cref{eqn:from semi-inclusive to inclusive}, and the soft coefficient $S_i$ for $i=5$ (as well as for $i=1,2,3$ \cite{Kretzer:1998ju}) is defined as
\begin{equation}
    S_i= \lim_{\hat{s}_{1} \rightarrow 0} \left( \frac{\Delta^2}{m_{2}^2 Q^2} \right)^{-\epsilon} \left( \frac{1}{\epsilon}\frac{\hat{s}_{1}^2}{m_{2}^2}\right) \int_{0}^{1} y^{-\epsilon} (1-y)^{-\epsilon} \left( \frac{m_{2}^2}{\hat{s}_1^2}+ \frac{m_{1}^2}{\hat{t}_{1}^2}+ \frac{\Sigma_{++}}{\hat{s}_1\hat{t}_1}       \right)  \, dy + \mathcal{O}(\epsilon)
\end{equation}
and $S_4=0$. 
Additionally, we set \(\mu^2 = Q^2\) \cite{Kretzer:1998ju} in both  \cref{eqn:vertex correction analitical expression} and \cref{eqn:phase space}, and following the \(\overline{\text{MS}}\) renormalization scheme \cite{Bardeen:1978yd}, absorb the Euler–Mascheroni constant along with the \(4\pi\) factor into the definition of \(\alpha_s\). Explicitly, the coefficients are given by
\begin{align}
    & S_4 =0, \nonumber\\
    & S_5= -\frac{1}{\epsilon}\left(-2+\Sigma_{++} I_1\right)+2+\frac{\Sigma_{++}}{\Delta}\left(\Delta I_1+\mathrm{Li}_2\left(\frac{2 \Delta}{\Delta-\Sigma_{++}}\right)-\mathrm{Li}_2\left(\frac{2 \Delta}{\Delta+\Sigma_{++}}\right)\right)\nonumber\\
    &\quad  +\ln \frac{\Delta^2}{m_2^2 Q^2}\left(-2+\Sigma_{++} I_1\right), \nonumber\\
    &  V_4= m_1 \left(  m_1 C_{q,+}+ \frac{q_-}{q_+} m_2 C_{q,-} \right), \nonumber\\
    &  V_5=   C_{R-}+\frac{1}{2}\Bigg((C_{1,+} +C_{q+})m_1^2+ C_{q-} m_2^2 \Bigg) + \frac{q_{-}}{q_+} \Bigg( \frac{1}{2} m_1 m_2 (C_{1,-}+ C_{q,-}+C_{q,+})+C_{+} \Bigg), \nonumber\\
    &N_4=  q_+ \frac{Q^2}{ \Delta}, \nonumber \\
    & N_5=  q_+ \frac{\Sigma_{+-}}{\Delta}.
\end{align}

\subsection{Wilson coefficients and NLO contribution}

To determine the contribution to the structure functions at NLO, it is crucial to first recall the general definition, expressed as a Mellin convolution product \cite{collins2004factorizationhardprocessesqcd}
\begin{equation}
 F_{i}^{\rm QS, \, NLO} =  \frac{\alpha_s}{2 \pi}  \mathcal{H}_{i}^{\rm QS, \, NLO} \otimes f,
 \label{eqn:main equation}
\end{equation}
where $f$ denotes the parton distribution function (PDF) and $\mathcal{H}_i$ are the inclusive Wilson coefficients, also referred to as the coefficient functions. The reason why $\alpha_s$ was introduced instead of $g_s^2$ lies in the fact that in the definition of the hadronic tensor the normalization factor involving $4\pi$ appears \cite{ParticleDataGroup:2024cfk}:
\begin{equation}
W_{\mu \nu} = \frac{1}{4 \pi} \int d^4 z \, e^{i q \cdot z} \langle P | \left[ J_\mu^{\dagger}(z), J_\nu(0) \right] | P\rangle\, .
\end{equation}
Following the same logic adopted for the partonic structure functions, we can define the semi-inclusive Wilson coefficients in the following way \cite{Kretzer:1998ju, Aivazis:1993kh,Aivazis:1993pi}:
\begin{equation}
 H_{i}^{\rm QS, \, NLO} \equiv \left(\frac{g_{s}^2}{2}\right)^{-1}  P_{i}^{\mu \nu} \overline{\hat{W}_{\mu \nu}^{\rm QS}} \Big|_{k_{1}^{+}= \xi P^{+}} , \quad i=4, \, 5
\label{eqn:mixing equation}
\end{equation}
with $P^2=M^2$. In this context, the notation indicates that the scalar products between the hadronic part, arising from the projectors, and the partonic part, originating from the partonic tensor, are determined by imposing proportionality between the plus components of the parton and the nucleon momenta. For a more detailed discussion, we refer to \cref{Appendix: Collinear Frame}.

By applying the hadronic projectors, it becomes clear that, in general, there is a mixing between the various partonic structure functions. Specifically, considering the Bjorken limit \cite{BjorkenLimit} (i.e.\ $Q^2 \gg M^2$),  we obtain the following relations:
\begin{align}
    \mathcal{H}_{4}^{\rm QS, \, NLO} &= C_F \left(\hat{\mathcal{F}}_4^{\rm QS, \, NLO} + (\mathcal{R}_2-1) \left( \mathcal{R}_1 \hat{\mathcal{F}}_2^{\rm QS, \, NLO} + \hat{\mathcal{F}}_5^{\rm QS, \, NLO} \right) \right)+\mathcal{O} \left( \frac{M^2}{Q^2} \right)\, ,
    \label{eqn: H4 mixing}\\
    \mathcal{H}_5^{\rm QS, \, NLO} &= C_F \mathcal{R}_2 \left( \hat{\mathcal{F}}_5^{\rm QS, \, NLO} + 2 \mathcal{R}_1 \hat{\mathcal{F}}_2^{\rm QS, \, NLO} \right)+\mathcal{O} \left( \frac{M^2}{Q^2} \right)\, ,
    \label{eqn:H5 mixing}
\end{align}
where
\begin{equation}
    \mathcal{R}_1 = \frac{2 m_{1}^2 \xi^\prime}{\Delta + \Sigma_{+-}}, \quad \mathcal{R}_2 = \frac{m_{1}^2 + Q^2 \mathcal{R}_1^2}{m_{1}^2 - Q^2 \mathcal{R}_1^2}\,.
    \label{eqn:coefficients mixing}
\end{equation}
Note that we have switched to the phase space integrated version of \cref{eqn:mixing equation}. Finally, the leading-order contribution is given by:

\begin{alignat}{4}
 & \mathcal{H}_{4}^{\mathrm{QS,\,LO}} &&= \mathcal{N}_{4}^{\rm QS, \, LO} \delta(1-\xi^\prime), \quad 
 &\quad &  F_{4}^{\rm QS, \, LO}(x,Q^2,m_1,m_2) &&= \mathcal{N}_{4}^{\rm QS, \, LO} f(\chi) \notag\\ 
 & \mathcal{H}_{5}^{\mathrm{QS,\,LO}} &&=  \mathcal{N}_{5}^{\rm QS, \, LO} \delta(1-\xi^\prime), \quad 
 &\quad &  F_{5}^{\rm QS, \, LO}(x,Q^2,m_1,m_2) &&= \mathcal{N}_{5}^{\rm QS, \, LO} f(\chi)
 \label{eqn:LO QS F4 F5}
\end{alignat}
where
\begin{align}
    &\mathcal{N}_{4}^{\rm QS, \, LO}=q_{+}  \frac{4 m_1^2 Q^2}   {\Delta}\frac{\left(\Sigma_{++}+\Delta\right)}{\left(\Sigma_{+-}+\Delta\right)^2}\, , \nonumber\\
    &\mathcal{N}_{5}^{\rm QS, \, LO}=q_{+}  \frac{m_1^2-m_2^2+\Delta}{Q^2}\, ,
    \label{eqn:normalization factors}
\end{align}
$f$ is the PDF of the incoming quark, and $\chi$ is defined in \cref{eqn: Chi definition}. The \( F_{4}^{\rm QS, \, LO } \) structure function vanishes in the massless limit, consistent with the Albright--Jarlskog relations \cite{ALBRIGHT1975467}.

Finally, the NLO contribution to the hadron level structure functions is given by:

\begin{equation}
   F_{i}^{\rm QS, \, NLO}(x,Q^2,m_1,m_2)=\frac{\alpha_s}{2 \pi } \int_\chi^1 \frac{d \xi^{\prime}}{\xi^{\prime}}\left[f\left(\frac{\chi}{\xi^{\prime}}, \mu^2\right) \mathcal{H}_i^{\rm QS, \, NLO}\left(\xi^{\prime},\frac{m_1}{Q}, \frac{m_2}{Q} \right)\right]\, .
   \label{eqn:NLO QS contribution}
\end{equation}
Throughout this work we use the notation \(\mathcal{H}(\xi', m_1/Q, m_2/Q)\) for the Wilson coefficients, rather than the more generic form \(\mathcal{H}(\xi', Q^2, m_1, m_2)\), as it makes their intrinsic scale invariance manifest. The same convention will be adopted later for the gluon–fusion channel.

\section{Boson-gluon fusion}
\label{sec:bgf}

For the boson-gluon fusion process, which appears for the first time at the next-to-leading order,\footnote{There is no process in the boson-gluon channel at $\mathcal{O}(\alpha_S^0)$, thus it only receives contributions from tree-like Feynman diagrams and there are no diagrams with loops.}
\begin{equation}
B^*(q) + g(p) \to \bar{Q}_1(k_1) + Q_2(k_2) ,
\end{equation}
the calculations are analogous to the real-emission calculation in the quark scattering case. The semi-inclusive partonic structure functions are defined by \cite{Kretzer:1998ju} 
\begin{equation}
\hat{F}_{i}^{\rm GF} \equiv \left(  \frac{g_{s}^2}{2 \pi} \right)^{-1} \hat{P}_{i, \, \rm GF}^{\mu \nu} \overline{\hat{W}_{\mu \nu}^{\rm GF}}, \quad i=4, \, 5,
\end{equation}
where the projectors $\hat{P}_{i, \, \rm GF}^{\mu \nu}$ are defined in \cref{eq:partonic projectors} and
\begin{equation}
    \overline{\hat{W}_{\mu \nu}^{\rm GF}} = \left( \frac{1}{8} \sum_{\rm color} \right) \left(\frac{1}{2} \sum_{\rm pol.} \right) \sum_{\rm spins}\hat{W}_{\mu \nu}^{\rm GF}.
    \label{eqn:average tensor boson gluon}
\end{equation}
In this expression, the factor of $1/8$ originates from the average over the color states of the incoming gluon, while the factor of $1/2$ accounts for averaging over its polarization states.

\begin{figure}[H]
    \centering
    \includegraphics[scale=0.8]{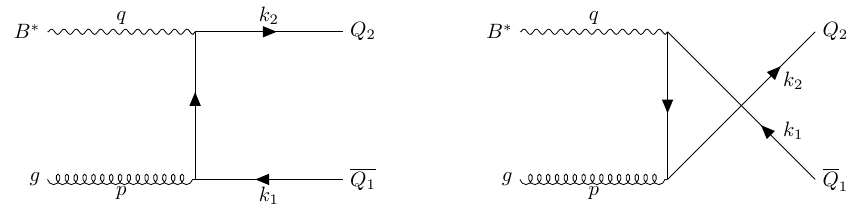}
    \caption{Feynman diagrams relevant for the real contribution at next-to-leading order to the boson-gluon fusion process.
    }
    \label{fig:Boson Gluon fusion}
\end{figure}

To determine the contribution of this process, we need to consider the relevant Feynman diagrams, as shown in \cref{fig:Boson Gluon fusion}. 
For convenience, the structure functions are expressed in terms of the variables $\zeta$ and $\hat{x}$, where
\begin{equation}
    \zeta = \frac{p \cdot k_2}{p \cdot q}, \quad \hat{s}=(q+p)^2=Q^2 \left( \frac{1}{\hat{x}}-1 \right).
\end{equation}
Additionally, the amplitude depends on $\hat{u} = (k_2 - p)^2$. The semi-inclusive partonic structure functions can then be decomposed as
\begin{equation}
    \hat{F}_{i}^{\rm GF} (\zeta, \hat{x}, Q^2)= 8 \pi \left(\frac{A_i(\hat{x})}{(1-\zeta)^2}   +\frac{B_i(\hat{x})}{\zeta^2}+\frac{C_i(\hat{x})}{1-\zeta}+\frac{D_i(\hat{x})}{\zeta}+E_i(\hat{x}) \right)
    \label{eq:BosonGluonFusionsemi-inclusive}
\end{equation}
where the functions \( A_i(\hat{x}) \), \( B_i(\hat{x}) \), \( C_i(\hat{x}) \), \( D_i(\hat{x}) \), and \( E_i(\hat{x}) \) are given by
\begin{align}
   & A_4= 0,   \nonumber\\ 
   &B_4=0,  \nonumber\\
   & C_4= -2 q_+ \frac{m_1^2}{Q^2} \hat{x}^2, \nonumber
   \\ & D_4= C_4 \,  (m_1 \leftrightarrow m_2), \nonumber \\
   &E_4=  -2 q_+ (\hat{x}-1) \hat{x}, \nonumber\\
   & A_5= q_+ \hat{x}^2 \frac{m_1^2}{Q^2} \left(1+\frac{ m_{2}^2-3 m_1^2}{Q^2} \right), \nonumber \\
    & B_5= A_5 (m_1 \leftrightarrow m_2), \nonumber\\
    & C_5=2 q_+\left(\frac{m_1^4 \hat{x}^2}{2 Q^4}-\frac{3 m_2^2 m_1^2 \hat{x}^2}{Q^4}+\frac{m_2^4 \hat{x}^2}{2
   Q^4}+\frac{m_1^2 \left(3-8 \hat{x}\right) \hat{x}}{2 Q^2}+\frac{m_2^2 \hat{x} \left(2
   \hat{x}-1\right)}{2 Q^2}+\frac{1}{4} \left(2 \left(\hat{x}-1\right) \hat{x}+1\right) \right), \nonumber \\
   & D_5= C_5 (m_1 \leftrightarrow m_2), \nonumber\\
    & E_5= -q_+(6 \hat{x}^2-6 \hat{x}+1).       
\end{align}
As for the quark scattering process, we integrate over the phase space with
\begin{equation}
    \hat{\mathcal{F}}_i^{\rm GF, \, NLO}(\hat{x},Q^2) \equiv  \frac{1}{8 \pi}\int_{\zeta_-}^{\zeta_+} \hat{F}_i^{\rm GF}(\zeta,\hat{x},Q^2) \, d \zeta,
\end{equation}
where the integration boundaries \( \zeta_{\pm} \) are defined as
\begin{equation}
    \begin{aligned}
        \zeta_{\pm} = \dfrac{1}{2}\left(\zeta_1 \pm \zeta_2\right) & \quad {\rm with} \quad
        \zeta_1 = 1 + \dfrac{m_2^2 - m_1^2}{\hat{s}}, \quad \zeta_2 = \dfrac{1}{\hat{s}}\Delta(\hat{s},m_2^2,m_1^2).
    \end{aligned}
\end{equation}
Using  \cref{eq:BosonGluonFusionsemi-inclusive}, we obtain
\begin{equation}
    \hat{\mathcal{F}}_i^{\rm GF, \, NLO}(\hat{x},Q^2) = A_i\dfrac{\zeta_+ - \zeta_-}{(1-\zeta_+)(1-\zeta_-)} + B_i\dfrac{\zeta_+ - \zeta_-}{\zeta_+\zeta_-} + C_i\ln\left(\dfrac{1-\zeta_-}{1-\zeta_+}\right) + D_i\ln\left(\dfrac{\zeta_+}{\zeta_-}\right) + E_i(\zeta_+ - \zeta_-).
    \label{WilsonGF}
\end{equation}


Unlike the quark scattering process, in this case there is no mixing, since the initial parton, the gluon, is massless. Therefore, we have
\begin{equation}
    \mathcal{H}_{i}^{\rm GF, \, NLO} = \hat{\mathcal{F}}_{i}^{\rm GF, \, NLO}.
\end{equation}
At the hadron level, the NLO contribution arising from the boson-gluon fusion process is then given by the convolution
\begin{equation}
    F_{i}^{\rm GF, \, NLO} = \frac{\alpha_s}{2\pi} \mathcal{H}_{i}^{\rm GF, \, NLO} \otimes g,
\end{equation}
where $g$ is the gluon PDF.  Explicitly, it takes the form
\begin{equation}
    F_{i}^{\rm GF, \, NLO}(x, Q^2, m_1, m_2) = \frac{\alpha_s}{2 \pi} \int_{\chi^\prime}^1 \frac{d \xi^{\prime}}{\xi^{\prime}} \left[ g\left(\frac{\chi^\prime}{\xi^{\prime}}, \mu^2\right) \mathcal{H}_i^{\rm GF, \, NLO} \left(\xi^{\prime},  \frac{m_1}{Q}, \frac{m_2}{Q}\right) \right],
    \label{boson gluon NLO}
\end{equation}
where
\begin{equation}
   \xi^\prime= \frac{\chi^\prime}{x} \hat{x} , \quad \chi^\prime = x \frac{(m_1 + m_2)^2 + Q^2}{Q^2}.
\end{equation}

\section{Structure functions in the ACOT scheme}
\label{sec:ACOT}

The treatment of heavy quarks in DIS requires a consistent framework that incorporates both mass effects near threshold and the resummation of large logarithms at asymptotically high scales. The ACOT scheme~\cite{Aivazis:1993kh,Aivazis:1993pi} provides such a framework by combining massive Wilson coefficients with DGLAP-evolved parton distribution functions, while introducing explicit subtraction terms to prevent double counting. This construction ensures that structure functions interpolate smoothly between the fixed-flavor-number scheme, where heavy quarks are produced dynamically, and the variable-flavor-number scheme, where heavy quarks are treated as active partons.

The conceptual basis of the ACOT scheme lies in the treatment of heavy quarks within the factorization framework \cite{Collins:1998rz}. In a massless scheme, collinear divergences in the partonic cross sections are regulated dimensionally and absorbed into the PDFs, so that heavy-quark parton distributions are generated dynamically through QCD evolution. When the heavy-quark mass is retained in the Wilson coefficients, it acts as a regulator of the collinear behavior, and heavy quarks are then produced explicitly in the hard scattering through processes such as boson--gluon fusion. The factorization scale $\mu$ controls the transition between these two descriptions: for $\mu \simeq m_1$, the mass-dependent treatment dominates, while for $\mu \gg m_1$, the heavy quark can be treated as a parton with its own distribution $f_{Q_1}(x,\mu^2)$.

Near threshold ($\mu \sim m_1$), the dominant contribution to the heavy-quark distribution arises from gluon splitting:
\begin{equation}
  f_{Q_1}(x,\mu^2) \simeq \frac{\alpha_s(\mu^2)}{2\pi} \ln\!\left(\frac{\mu^2}{m_1^2}\right)
  \, P_{qg} \otimes g(x,\mu^2) \;+\; \mathcal{O}(\alpha_s^2).
\end{equation}
When this PDF is convolved with the LO quark-scattering coefficient function, a term of apparent $\mathcal{O}(\alpha_s)$ in the structure functions actually appears at $\mathcal{O}(\alpha_s^2)$ (since $f_{Q_1} = \mathcal{O}(\alpha_s)$):
\begin{equation}
  F_i^H \simeq \frac{\alpha_s(Q^2)}{2\pi} \ln\!\left(\frac{Q^2}{m_1^2}\right)
  \Big[ \mathcal{H}_i^{\rm QS,LO} \otimes_\chi P_{qq} \otimes f_{Q_1}
      + \mathcal{H}_i^{\rm QS,LO} \otimes_\chi P_{qg} \otimes g \Big],
  \quad i=4,5,
\end{equation}
where $P_{qq}$ and $P_{qg}$ are the usual leading order splitting functions \cite{Altarelli:1977zs} defined as
\begin{equation}
    P_{qq}(\xi^\prime)=  C_F \left[\frac{1+\xi^{\prime \, 2}}{1-\xi^\prime} \right]_+ , \quad P_{qg}(\xi^\prime)=  \frac{1}{2} \left[(1-\xi^\prime)^2+\xi^\prime\right]
\end{equation}
with the standard convolution \cite{Collins:1989gx}
\begin{equation}
    (P \otimes f)(x) = \int_x^1 \frac{d\xi'}{\xi'}\, P(\xi')\, f\!\left(\frac{x}{\xi'}\right).
\end{equation}

At genuine NLO, the massive quark-scattering and gluon-fusion coefficient functions develop exactly the same large logarithms in the limit $m_{Q_1}\to 0$:

\begin{align}
  \lim_{m_1\to 0} \mathcal{H}_i^{\rm QS,NLO} \otimes_\chi f_{Q_1}
  &= \frac{\alpha_s}{2\pi} \ln\!\left(\frac{Q^2}{m_{1}^2}\right) P_{qq} \otimes f_{Q_1}
     + \text{finite}, \\
  \lim_{m_1\to 0} \mathcal{H}_i^{\rm GF,NLO} \otimes g
  &= \frac{\alpha_s}{2\pi} \ln\!\left(\frac{Q^2}{m_1^2}\right) P_{qg} \otimes g
     + \text{finite}.
\end{align}
A direct addition of these NLO terms to the LO massive calculation would therefore double-count the resummed logarithms already contained in the heavy-quark PDF.

The ACOT scheme resolves this by subtracting those contributions, that are reproduced by the zero-mass (ZM) limit, from the massive NLO coefficient functions. Precisely we have:
\begin{equation}
  F_i^{\rm ACOT} = F_i^{\rm QS,LO} + F_i^{\rm QS,NLO} + F_i^{\rm GF,NLO}
  - F_i^{\rm QS,SUB} - F_i^{\rm GF,SUB},
  \qquad i=4,5.
  \label{eq:ACOT_contribution_definition}
\end{equation}

In the following, we present the explicit subtraction terms adopting the conventional scale choice $\mu_R = \mu_F = Q$, as originally derived in the ACOT scheme~\cite{Aivazis:1993pi, Collins:1998rz, Kretzer:1998ju}. This choice serves to simplify the analytic structure of the coefficients by eliminating explicit renormalization scale logarithms in the definition.  However, we emphasize that this choice is not mandatory. In modern phenomenological applications and global PDF analyses, more general dynamical scales such as $\mu_{R,F}^2 = Q^2 + c m_Q^2$ or scales proportional to the transverse mass are commonly employed to improve perturbative stability near thresholds and in multi-scale kinematic regions. With this in mind, the subtraction terms for the inclusive case are given by \cite{Kretzer:1998ju, 1991NuPhB.361..626M}
\begin{align}
  F_i^{\rm QS,SUB}\left(x, Q^2, m_1, m_2\right) &= \mathcal{N}_i^{\rm QS,LO} \frac{\alpha_s(Q^2)}{2\pi}
  \int_{\chi}^1 \frac{\mathrm{d}\xi'}{\xi'} \,
  f_{Q_1}\!\left(\frac{\chi}{\xi'},Q^2\right) \,
  \mathcal{H}_i^{\rm QS,SUB}\left(\xi', \frac{m_1}{Q}\right), \\[4pt]
  \mathcal{H}_i^{\rm QS,SUB}\left(\xi', \frac{m_1}{Q}\right) &= c_i C_F \Bigg[ \frac{1+\xi'^2}{1-\xi'}
  \left( \ln\frac{Q^2}{m_1^2} - 1 - 2\ln(1-\xi') \right) \Bigg]_+ ,
  \qquad (c_4=0,\; c_5=1),
\end{align}

and for the gluon-fusion channel
\begin{align}
  F_i^{\rm GF,SUB}\left(x, Q^2,  m_1, m_{2}\right) &= \mathcal{N}_i^{\rm QS,LO} \frac{\alpha_s(Q^2)}{2\pi}
  \int_{\chi}^1 \frac{\mathrm{d}\xi'}{\xi'} \,
  g\!\left(\frac{\chi}{\xi'},Q^2\right) \,
  \mathcal{H}_i^{\rm GF,SUB}\left(\xi', \frac{m_1}{Q}, \frac{m_2}{Q}\right), \\[4pt]
  \mathcal{H}_i^{\rm GF,SUB}\left(\xi', \frac{m_1}{Q}, \frac{m_2}{Q}\right) &= c_i P_{q g}(\xi^\prime)\left[ \Theta \left(1-\frac{m_1^2}{Q^2}\right) \ln\frac{Q^2}{m_1^2}
  + \Theta \left(1-\frac{m_2^2}{Q^2}\right) \ln\frac{Q^2}{m_2^2} \right].
  \label{GF subtraction terms}
\end{align}
Here, the Heaviside theta-functions enforce the CWZ decoupling prescription \cite{PhysRevD.18.242}: the subtraction is only active when the factorization scale $Q$ exceeds the heavy quark mass. While the first term subtracts the PDF double-counting associated with the incoming parton $m_1$, the inclusion of the $m_2$ term is necessary to explicitly subtract the collinear singularity arising from the second heavy quark line, ensuring the result is IR safe. 
Below threshold, the heavy quark is not considered an active parton, and no collinear subtraction is required. The ACOT-renormalized Wilson coefficients are then given by:

\begin{align}
  \mathcal{H}_i^{\rm QS,ACOT}\left(\xi', \frac{m_1}{Q}, \frac{m_2}{Q}\right) &=
  \mathcal{H}_i^{\rm QS,NLO}\left(\xi', \frac{m_1}{Q}, \frac{m_2}{Q}\right) - \mathcal{H}_i^{\rm QS,SUB}\left( \xi^\prime, \frac{m_1}{Q}\right), \\[4pt]
  \mathcal{H}_i^{\rm GF,ACOT}\left(\xi', \frac{m_1}{Q}, \frac{m_2}{Q}\right) &=
  \mathcal{H}_i^{\rm GF,NLO}\left(\xi', \frac{m_1}{Q}, \frac{m_2}{Q}\right) - \mathcal{H}_i^{\rm GF,SUB}\left(\xi', \frac{m_1}{Q}, \frac{m_2}{Q}\right).
\end{align}
By construction, these coefficient functions are free of large logarithms and smoothly approach the standard massless $\overline{\rm MS}$ NLO coefficient functions in the double limit $m_1,m_2\to 0$:
\begin{align}
  \lim_{m_2\to 0}\lim_{m_1\to 0} \mathcal{H}_i^{\rm QS,ACOT}\left(\xi', \frac{m_1}{Q}, \frac{m_2}{Q}\right) &=
  C_{F,i}^{(1)}(\xi'), \\[4pt]
  \lim_{m_2\to 0}\lim_{m_1\to 0} \mathcal{H}_i^{\rm GF,ACOT}\left(\xi', \frac{m_1}{Q}, \frac{m_2}{Q}\right) &=
  C_{G,i}^{(1)}(\xi').
  \label{zero mass coefficients}
\end{align}
The coefficient functions $C_{F/G, i}^{(1)}(\xi^\prime)$ were computed previously in Ref.~\cite{Kretzer_2002} (eqs.~(31)--(34)) and are reproduced here for completeness:

\begin{align}
C_{G,4}^{(1)}(\xi') &= 2\,\xi'(1 - \xi'), \label{Coefficient Functions First}\\[4pt]
C_{G,5}^{(1)}(\xi') &= -1
  + 8\,\xi'(1 - \xi')
  + \big[(1 - \xi')^{2} + \xi'^{2}\big]\,
    \ln\!\left(\frac{1 - \xi'}{\xi'}\right), \\[6pt]
C_{F,4}^{(1)}(\xi') &= C_F\,\xi', \\[4pt]
C_{F,5}^{(1)}(\xi') &= 
  C_F \left[
     \frac{1 + \xi'^2}{1 - \xi'} 
     \left(
       \ln\!\left(\frac{1 - \xi'}{\xi'}\right)
       - \frac{3}{4}
     \right)
     + \frac{1}{4}(9 + 5\xi')
  \right]_{+}.
  \label{Coefficient Functions Last}
\end{align}

An immediate consequence of the previous equations is that in the ultrarelativistic
($m_f^2/Q^2 \to 0$) limit, all explicit mass dependence cancels.  
In this regime the structure functions satisfy the
Albright$-$Jarlskog relations \cite{ALBRIGHT1975467},
\begin{equation}
  F_L^{\rm ACOT}(x,Q^2) = 2x\,F_4^{\rm ACOT}(x,Q^2),
  \qquad
  F_2^{\rm ACOT}(x,Q^2) = x\,F_5^{\rm ACOT}(x,Q^2),
  \label{eqn:AJ_massless}
\end{equation}
with  $F_L^{\rm ACOT}(x,Q^2)= F_2^{\rm ACOT}(x,Q^2)- 2x F_1^{\rm ACOT}(x,Q^2)$. This relation holds for both neutral-current and charged-current scattering.

In summary, the ACOT scheme provides a theoretically consistent and phenomenologically robust description of the heavy-quark structure functions. It retains exact heavy-quark mass dependence at all orders in the hard-scattering coefficient functions, avoids double counting of logarithmic contributions, and automatically resums heavy quarks collinear logs to all orders via DGLAP evolution. The explicit expressions presented above serve as the foundation for the numerical analysis performed in the following section.\\

\section{Numerical implementation and results}
\label{sec:numerical}

In this section, we discuss explicit numerical results that illustrate important features of the structure functions $F_4$ and $F_5$ as well as their relevant kinematics limits of the ACOT scheme. We have implemented NC interactions in \texttt{Mathematica} and further the CC interactions for both $W^-$ and $W^+$ exchange in \texttt{Mathematica} and in the \texttt{C++} library \APFEL{}~\cite{Bertone:2013vaa,Bertone:2017gds,APFELppGithub}. \APFEL{} provides a highly optimized setup for the calculation of structure functions by the means of constructing pre-computed tables that allow for rapid numerical evaluations across a wide range of kinematics\footnote{We did not implement the NC structure functions in \APFEL{} due to the unusual and phenomenologically less relevant case that only $Z$-boson exchanges contribute which would require a substantial reorganization of the existing framework and is not justified within the scope of the present study.}.

The implementation of the massive $F_4$ and $F_5$ structure functions in the ACOT scheme has to be performed on a channel-by-channel basis due to the different flavor and mass dependences in the NC and CC interactions. For the following it is useful to introduce the short-hand notation 
\begin{equation}
  \mathcal{H}_{i,\alpha,\beta}^{\mathrm{QS/GF},k}(\xi')
  \equiv \mathcal{H}_{i}^{\mathrm{QS/GF}}
  \!\left(\xi',\frac{m_\alpha}{Q},\frac{m_\beta}{Q}\right),
  \label{eqn:index_convention_coefficients}
\end{equation}
where $\alpha,\beta\in\{u,d,s,c,b,t\}$ and $k=ZZ,WW$ in order to clarify the mass dependence for the individual flavor contributions.
The antiparticles are denoted by $\bar\alpha$ (with $m_\alpha=m_{\bar\alpha})$  
and the couplings are given by \cite{HalzenMartin1984}
\begin{align}
  ZZ\text{, up-type case}:\quad & V=V'=\phantom{-}\tfrac12-\tfrac43\sin^2\theta_W,\;
  A=A'=\phantom{-}\tfrac12, \\[4pt]
  ZZ\text{, down-type case}:\quad & V=V'=-\tfrac12+\tfrac23\sin^2\theta_W,\;
  A=A'=-\tfrac12, \\[4pt]
  WW:\quad & V=V'=A=A'=1. 
\end{align}
Following \cref{eq:ACOT_contribution_definition}, the individual massive contributions at LO, NLO and the subtraction terms are combined into the ACOT scheme as
\begin{align}
  F_{i,\alpha,\beta}^{\mathrm{QS},\mathrm{ACOT},k}(x,Q^2) &=
  F_{i,\alpha,\beta}^{\mathrm{QS,LO},k}
  +F_{i,\alpha,\beta}^{\mathrm{QS,NLO},k}
       -F_{i,\alpha,\beta}^{\mathrm{QS,SUB},k},
  \\[4pt]
  F_{i,\alpha,\beta}^{\mathrm{GF},\mathrm{ACOT},k}(x,Q^2) &=\,\,\phantom{F_{i,\alpha,\beta}^{\mathrm{QS,LO},k}
  + }F_{i,\alpha,\beta}^{\mathrm{GF,NLO},k}
  -F_{i,\alpha,\beta}^{\mathrm{GF,SUB},k}.
\end{align}
Then the full neutral-current structure functions are 
\begin{align}
  F_i^{\mathrm{NC,ACOT}}(x,Q^2) &=
  \sum_{U=u,c,t}
  \Bigl[\,
    F_{i,U,U}^{\mathrm{QS,ACOT},ZZ}
  + F_{i,\bar U,\bar U}^{\mathrm{QS,ACOT},ZZ}
  + 2\,F_{i,U,U}^{\mathrm{GF,ACOT},ZZ}
  \,\Bigr] \nonumber\\
  &+ \sum_{D=d,s,b}
  \Bigl[\,
    F_{i,D,D}^{\mathrm{QS,ACOT},ZZ}
  + F_{i,\bar D,\bar D}^{\mathrm{QS,ACOT},ZZ}
  + 2\,F_{i,D,D}^{\mathrm{GF,ACOT},ZZ}
  \,\Bigr].
  \label{eqn:NC_ACOT_full}
\end{align}
The factor of two in front of the gluon-fusion terms arises from the symmetry $F_{i,\bar f,\bar f}^{\mathrm{GF,ACOT},ZZ}= F_{i,f,f}^{\mathrm{GF,ACOT},ZZ}$ for any flavor $f$.
For the charged-current case the corresponding expressions are
\begin{align}
  F_i^{W^-,\mathrm{ACOT}}(x,Q^2) =
  \sum_{U=u,c,t}\sum_{D=d,s,b}|V_{UD}|^2
  \Bigl[
    &\phantom{+}\;\,F_{i,\bar D,\bar U}^{\mathrm{QS,ACOT},WW}
  + F_{i,U,D}^{\mathrm{QS,ACOT},WW}\nonumber\\
  &+ F_{i,\bar D,\bar U}^{\mathrm{GF,ACOT},WW}
  + F_{i,U,D}^{\mathrm{GF,ACOT},WW}
  \,\Bigr],
  \nonumber\\[4pt]
  F_i^{W^+,\mathrm{ACOT}}(x,Q^2) =
  \sum_{U=u,c,t}\sum_{D=d,s,b}|V_{UD}|^2
  \Bigl[
    &\phantom{+}\;\,F_{i,D,U}^{\mathrm{QS,ACOT},WW}
  + F_{i,\bar U,\bar D}^{\mathrm{QS,ACOT},WW}\nonumber\\
  &+ F_{i,D,U}^{\mathrm{GF,ACOT},WW}
  + F_{i,\bar U,\bar D}^{\mathrm{GF,ACOT},WW}
  \,\Bigr].
\end{align}

The Wilson coefficients for $F_4$ and $F_5$, especially when retaining the quark masses, are not straightforward to evaluate numerically.
Since the coefficients not only consist of regular functions, but also of delta- and plus-distributions, a special treatment is required. In the case of plus-distributions, the integration is further complicated, since the integration is not performed over the interval $[0,1]$ but $[\chi,1]$\footnote{Note that up to NLO plus-distributions only appear in the QS case for $F_{4/5}$, therefore the relevant lower integration limit here is $\chi$.}. This leads to additional terms in the evaluation of the integrals, see \cref{appendix: Plus distribution identities} for a derivation. The numerical treatment along with explicit calculations of the additional terms for $F_4$ and $F_5$ can be found in \cref{appendix: Plus Distribution Treatment}. 

By examining the {\em unpolarized} NC cross section, \textit{cf.}~\cref{eq:ZZ cross section}, 
we can see that the contributions from $F_4$ and $F_5$ are suppressed by both the lepton mass and the $Z$ propagator
such that the cross section is dominated by the structure functions $F_{1,2,3}$ which
also receive contributions from the $\gamma$ and $\gamma Z$ channels.
Therefore we discuss the NC results only briefly in the following and focus more on the CC results. 

All numerical evaluations use the \texttt{CT18NLO} PDF set~\cite{Hou:2019efy} accessed via LHAPDF~\cite{Buckley:2014ana} using the masses $m_c=1.3\,\mathrm{GeV}$ and $m_b=4.75\,\mathrm{GeV}$. Contributions from the top quark are neglected in the plots.

\subsection{Neutral-current results}

\begin{figure}[]
    \centering
    \includegraphics[width=0.5\linewidth]{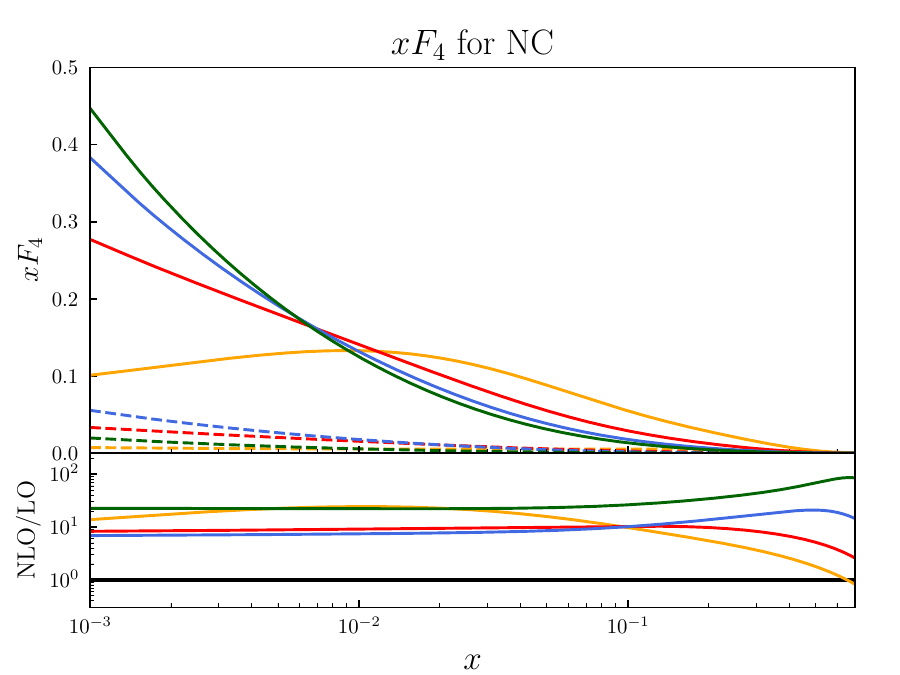}%
    \includegraphics[width=0.5\linewidth]{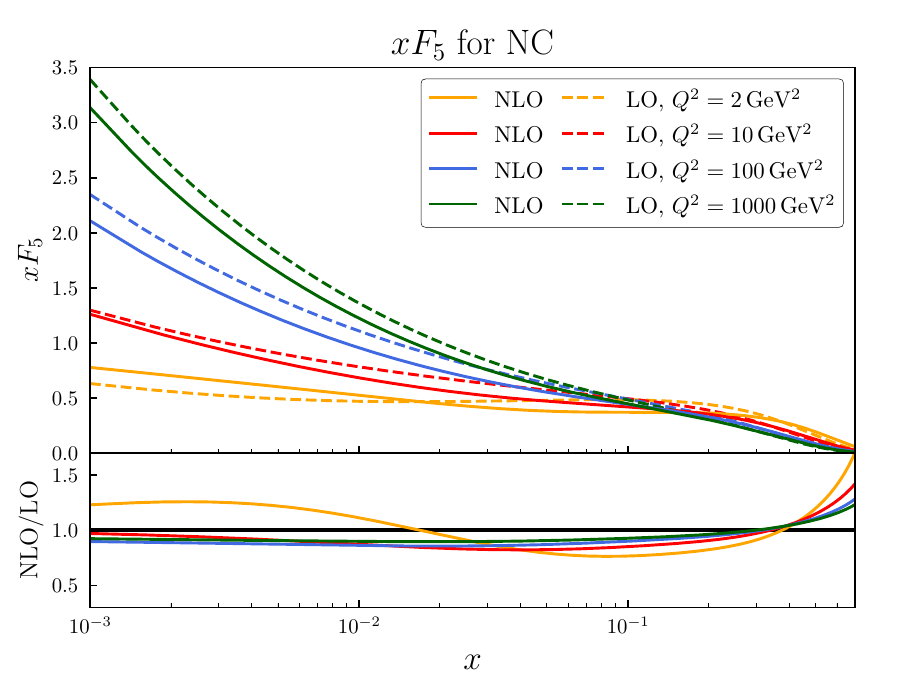}
    \caption{The total structure functions $xF_4$ (left) and $xF_5$ (right) for $Z$ exchange at NLO for various $Q^2$ values. The LO contributions are given by the dashed lines. The bottom panel shows the NLO over LO ratio on a logarithmic scale for $F_4$ and on a linear scale for $F_5$. Note the difference in the scales between $F_4$ and $F_5$ for both the absolute and ratio plots.}
    \label{fig:NCF4F5_NLO_LO}
\end{figure}

We show calculations of the absolute structure functions in \cref{fig:NCF4F5_NLO_LO} with $xF_{4}$ on the left and $xF_{5}$ on the right in the range $x\in[10^{-3},0.7]$. The upper panels show the NLO (solid) and LO (dashed) curves for $Q^2\in\{2,10,100,1\,000\}$ GeV${}^2$ in the colors \{yellow, red, blue, green\}. Note that we choose to plot the structure function multiplied with $x$, since both are accompanied by $x$ in the cross section. The lower panels depict the NLO/LO ratios. 
We can see that, both in terms of the absolute scale and the size of the NLO/LO correction, $F_4$ behaves similar to $F_L$ and $F_5$ similar to $F_2$, as indicated by the \AJrelation{} (\textit{cf.}~\cref{eqn:AJ_massless}). The shape of $F_4$ is mainly driven by the gluon contribution at NLO. The LO contribution is suppressed by the quark masses and effectively zero at large $Q^2$. 
On the other hand, $F_5$ receives only moderate corrections at NLO, and one can see the distinct valence-quark shape at larger $x$ arising from the LO. The NLO corrections can be positive or negative depending on $x$ and $Q$. The corrections go up to $+25\%$ at lower $x$ and $Q$. Close to $x=1$ the relative corrections become very large, where the perturbative series becomes unstable \cite{Catani:1989ne, Moch_2009} due to $\ln(1-\chi)$ terms (see \cref{appendix: Plus Distribution Treatment} for explicit appearances in $F_5$). However, the absolute corrections remain small as the structure function strongly tends to zero. 

\subsection{Charged-current results}

\begin{figure}[t]
    \centering
    \includegraphics[width=0.5\linewidth]{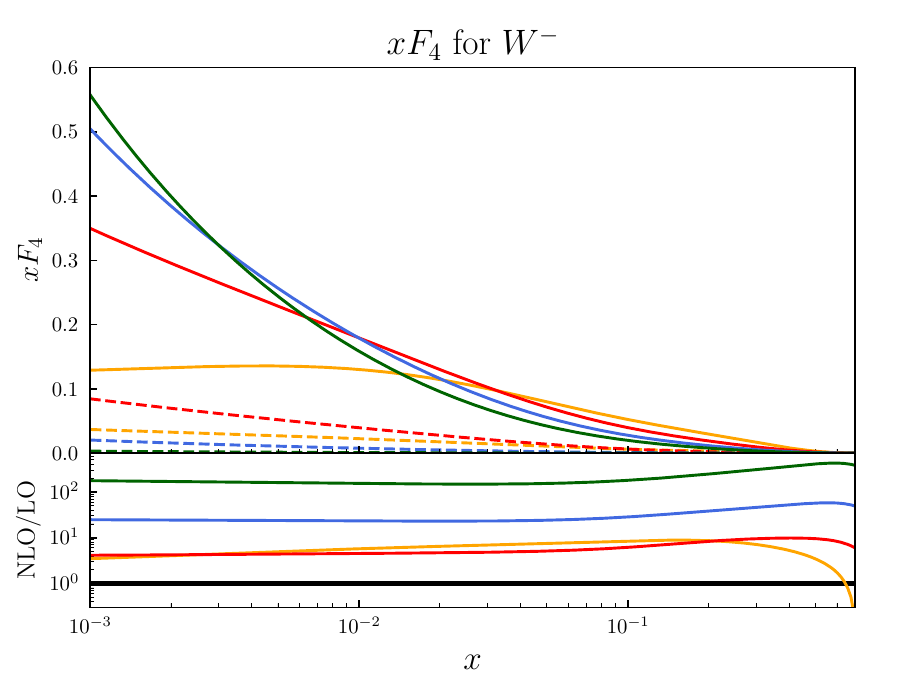}%
    \includegraphics[width=0.5\linewidth]{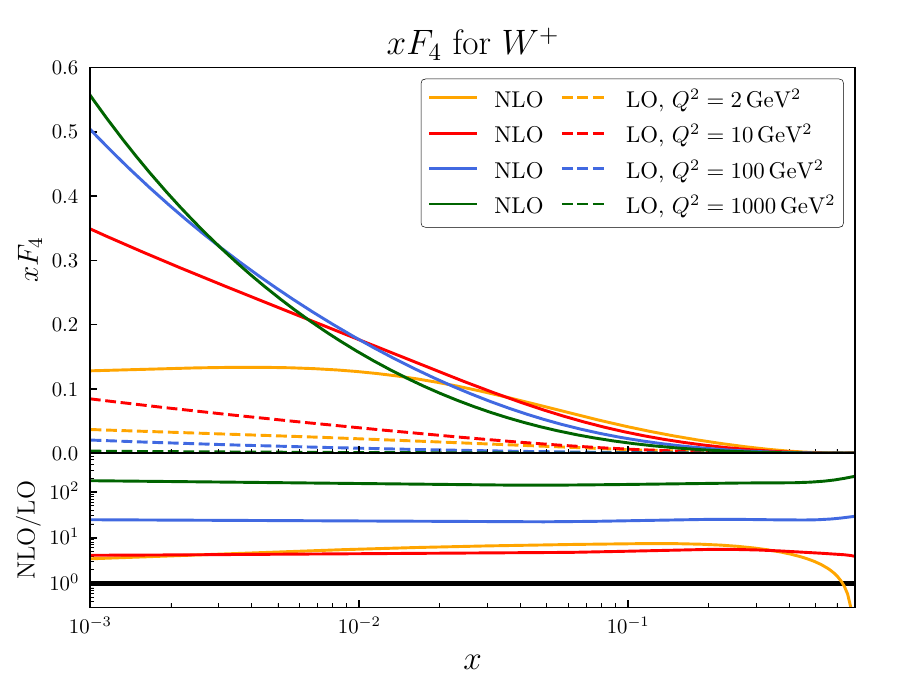}
    \includegraphics[width=0.5\linewidth]{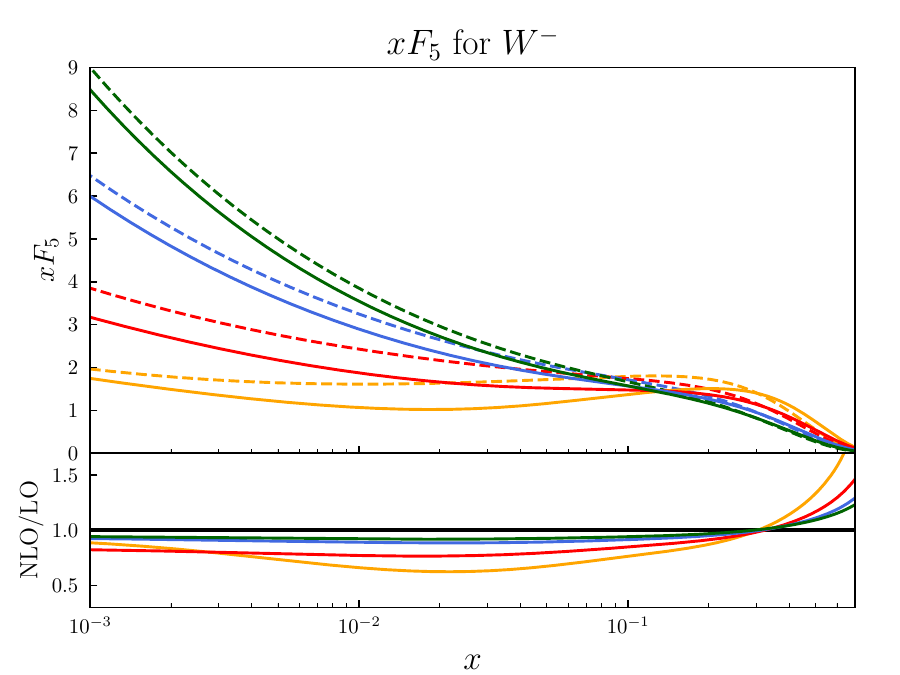}%
    \includegraphics[width=0.5\linewidth]{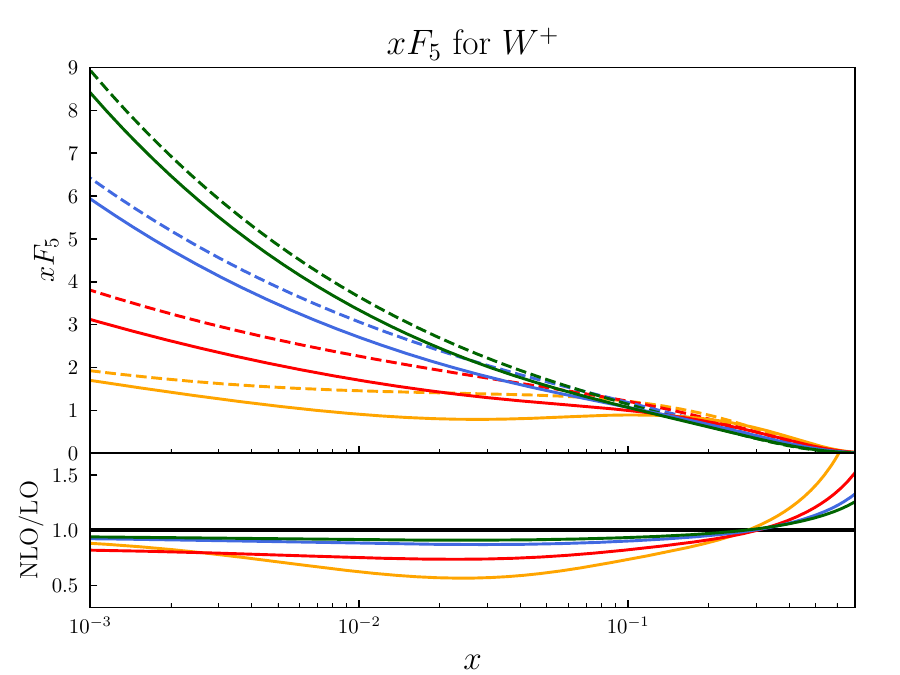}
    \caption{Same as \cref{fig:NCF4F5_NLO_LO} but for CC. We show $xF_4$ in the top row and $xF_5$ in the bottom row for $W^-$ exchange in the left column and $W^+$ exchange in the right column. }
    \label{fig:CCF4F5_NLO_LO}
\end{figure}

The CC structure functions given in the same format as for the NC case are shown in \cref{fig:CCF4F5_NLO_LO}. The top row shows $xF_4$, and the bottom row $xF_5$. On the left hand side (here and in the following figures) we assume $W^-$ exchange and $W^+$ exchange on the right hand side. Comparing this figure to the NC results, we note generally the same behavior: $F_4$ is dominated by the gluon contribution and receives large NLO corrections, whilst $F_5$ is mainly given by the quarks from the leading order QS. For the latter in CC the NLO corrections can be larger and yield up to $\sim45\%$. If we compare the predictions for $W^-$ and $W^+$ exchanges, we observe differences originating from the flavor structure. The GF contribution is the same for $W^-$ and $W^+$ and therefore we find the strongest differences for $F_5$. We see that the valence bump is more pronounced for $W^-$, since it probes the up-quark and $W^+$ the down-quark. Here, we have chosen the proton as the hadron. The difference would be reversed, if we were to choose a neutron as the target, and reduced (or completely removed) for (isoscalar) nuclei. In the sea-quark region, at lower $x$, we find the same $F_5$ for both $W^-$ and $W^+$ as the difference between up- and down-type quarks vanishes. 

\begin{figure}[t]
    \centering
    \includegraphics[width=0.5\linewidth]{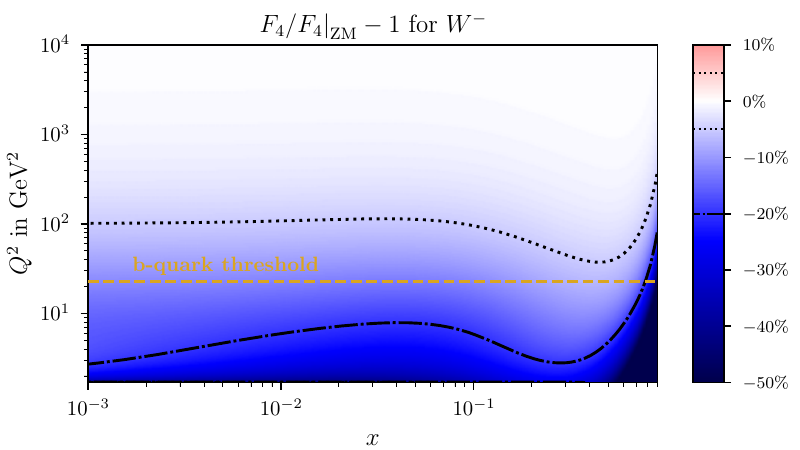}%
    \includegraphics[width=0.5\linewidth]{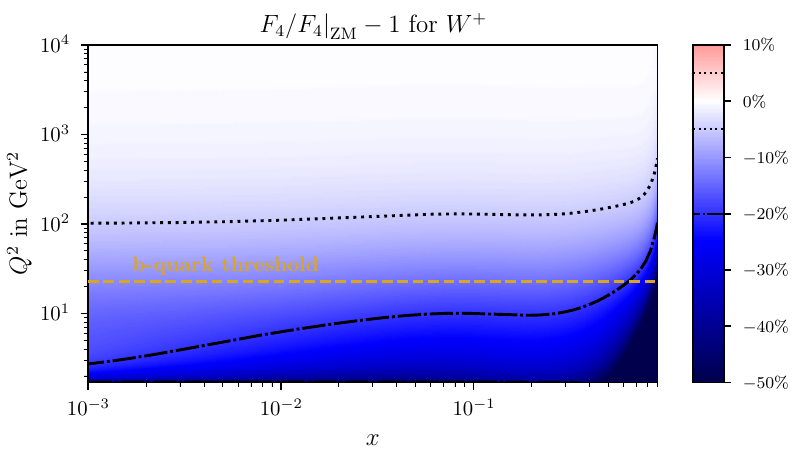}
    \includegraphics[width=0.5\linewidth]{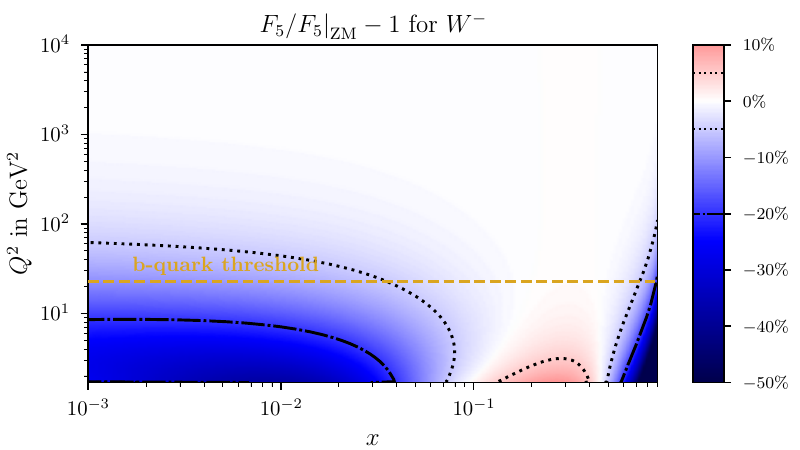}%
    \includegraphics[width=0.5\linewidth]{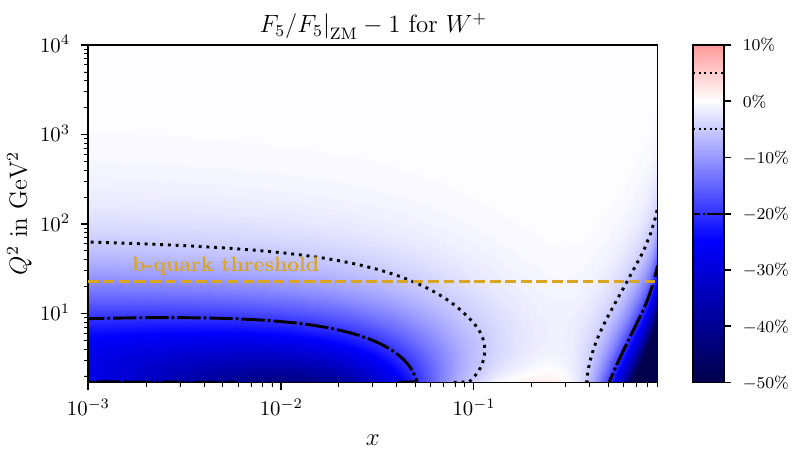}
    \caption{The ratios of \cref{eq:CC_ratio_ACOT_ZM} at NLO for $F_4$ in the top row and $F_5$ in the bottom row. The dash-dotted line indicates the -20\% difference and the dotted line the -5\% difference. Note that all four ratios reach -100\% in the bottom right corner.}
    \label{fig:CCF4F5_fullACOT_over_ZM}
\end{figure}

To illustrate the mass dependence of our results we now turn to ratios to the zero-mass expressions for both structure functions. Due to the \AJrelation{}, with $\left.2xF_4\right|_{ZM}=\left.F_L\right|_{ZM}$ and $\left.xF_5\right|_{ZM}=\left.F_2\right|_{ZM}$, these ratios also show the convergence towards the more familiar $F_{L/2}$ structure functions. In \cref{fig:CCF4F5_fullACOT_over_ZM} we display the ratios
\begin{alignat}{3}
\label{eq:CC_ratio_ACOT_ZM}
    \frac{F^{W^-}_{4}(x,Q^2)}{\left.F^{W^-}_{4}(x,Q^2)\right|_{\rm ZM}}-1 \qquad &\text{and}&\qquad \frac{F^{W^+}_{4}(x,Q^2)}{\left.F^{W^+}_{4}(x,Q^2)\right|_{\rm ZM}}&-1 \\
    \frac{F^{W^-}_{5}(x,Q^2)}{\left.F^{W^-}_{5}(x,Q^2)\right|_{\rm ZM}}-1 \qquad &\text{and}&\qquad \frac{F^{W^+}_{5}(x,Q^2)}{\left.F^{W^+}_{5}(x,Q^2)\right|_{\rm ZM}}&-1
\end{alignat}
as heatmaps. These figures are arranged in the same order as the ratios above. We consider the ratio on a double-logarithmic grid of $(400 \times 400)$ points corresponding to the ranges $Q^2 \in [m_c^2,10^4]$ GeV${}^2$ and $x \in \left[10^{-4} , 0.9\right]$. The lower limit of the $Q^2$ range is given by the initial scale of the \texttt{CT18NLO} PDFs. Deviations from zero are given in blue for negative values and red for positive values. The dashed-dotted contour indicates $-20\%$ difference and the dotted $\pm5\%$. Note that the color coding for the heatmap is kept the same for all four panels.

The first row, showing $F_4$, yields almost uniform suppression across all values of $x$ of the massive calculation compared to the ZM. The mass effects can be very significant at low $Q^2$, especially in the bottom-right (high-$x$) corner, where the ratios almost reach $-100\%$, but diminish quickly. The $-5\%$ contour is already found at $Q^2\approx10^2\,\rm{GeV}^2$ and the $-1\%$ contours (not shown in the figure) are at $Q^2\approx10^3\,\rm{GeV}^2$. The only variation is in the valence region for $W^-$, where the contour dips towards lower $Q^2$ values. Since $F_4$ is, for the most part, given by the GF contribution, a large part of the suppression is easily identified as the mass-reduced phase space given in the equations through the lower integration bound $\chi^\prime$.
The second row, showing $F_5$, yields a different behavior along the $x$ axes. For low- and very high-$x$ we find a similar suppression as for $F_4$, but in the valence region the suppression fades, or, in the case for $W^-$, even becomes an $+5\%$ enhancement. Overall the mass effect die out faster as for $F_4$, with the $-5\%$ contour reaching up to $80\,\rm{GeV}^2$ and the $-1\%$ contour (not shown) reaching up to $500\,\rm{GeV}^2$.

\section{Conclusions}
\label{sec:conclusions}

In this work, we have presented a comprehensive analysis of the deep-inelastic structure functions $F_4$ and $F_5$ at NLO in QCD. Our framework incorporated full heavy quark mass effects within the ACOT scheme. We provided both analytic expressions and a detailed numerical implementation, accounting for quark scattering and boson-gluon fusion contributions with full real and virtual corrections.

Our analysis revealed that NLO corrections lead to substantial modifications of leading-order predictions for $F_5$, with deviations reaching $25\text{--}45\%$ in relevant kinematic regions. In contrast, $F_4$ is almost entirely generated at NLO, as its leading order contribution is strongly suppressed in the massless limit. These effects are particularly pronounced at low $Q^2$ and moderate $x$, where heavy quark mass effects are significant. Specifically, we confirm that $F_4$ remains suppressed by the quark mass, whereas $F_5$ exhibits nontrivial behavior, particularly in the presence of heavy flavors. In the large $Q^2$ region (above $Q^2=10^3\,\rm{GeV}^2$ for less than 1\% correction) we confirm the onset of the \AJrelation{}, which yields a connection between $F_{4/5}$ and the better known $F_{L/2}$. 

The theoretical framework and numerical results presented here are directly relevant for upcoming experimental programs, such as the SHiP experiment at CERN and neutrino telescopes like IceCube. These facilities are uniquely sensitive to contributions from $F_4$ and $F_5$ via $\tau$-lepton and $\nu_\tau$ interactions. Consequently, this work establishes a baseline for interpreting future measurements and probing the heavy quark content of the nucleon with unprecedented precision.

Future investigations should address the interplay between NLO corrections, target mass effects, and nuclear corrections to achieve a fully consistent description of DIS structure functions across the entire kinematic range. Furthermore, a quantitative assessment of theoretical uncertainties and an extension to next-to-next-to-leading order will be crucial for maximizing the impact of future experimental data.

\acknowledgments

This work has been supported by the BMBF under contract 05P24PMA. M.K.\ thanks the School of Physics at the University of New South Wales in Sydney, Australia for its hospitality and financial support through the Gordon Godfrey visitors program. 
The work of P.R.~was supported by the U.S.~Department of Energy under Grant \mbox{No.~DE-SC0010129} 
and by the Office of Science, the Office of Nuclear Physics, within the framework of the Saturated Glue (SURGE) Topical Theory Collaboration.
P.R. further thanks the Jefferson Lab for their hospitality. This material is based upon work supported by the U.S. Department of Energy, Office of Science, Office of Nuclear Physics under contract DE-AC05-06OR23177.

\appendix

\section{Collinear frame } \label{Appendix: Collinear Frame}

When investigating the scattering of a space-like vector boson off a nucleon, it is convenient to adopt reference frames where the relevant four-vectors, \( P \) and \( q \), lie in the \( t \)-\( z \) plane. These frames, often referred to as collinear frames, simplify the analysis by aligning the motion along the longitudinal axis. In such frames, it is particularly useful to express the four-vectors using their light-cone coordinate components \cite{LightConeDirac}. In the collinear frame, a general four-vector \( v^\mu \) can then be written as
\begin{equation}
    v^\mu = (v^+, v^-, \boldsymbol{v_\bot}),
\end{equation}
where the "plus" and "minus" components are defined as
\begin{equation}
    v^\pm = \frac{1}{\sqrt{2}}(v^0 \pm v^3)
\end{equation}
and the transverse components are given by \( \boldsymbol{v_\bot} = (v^1, v^2) \). In this coordinate system, the invariant line element becomes
\begin{equation}
    ds^2 = 2 dx^+ dx^- - \delta_{ij} dx^i dx^j \quad \text{with} \quad i,j = 1,2.
    \label{Collinear frame metric}
\end{equation}
This metric reflects the separation of longitudinal and transverse components, which is a key feature of light-cone coordinates.

In light-cone coordinates, the four-momenta \( P^\mu \) and \( q^\mu \) can be expressed as \cite{ELLIS198329}
\begin{align}
    P^\mu &= \left( P^+, \frac{M^2}{2P^+}, \boldsymbol{0} \right), \label{Collinear frame P} \\
    q^\mu &= \left( -\kappa P^+, \frac{Q^2}{2\kappa P^+}, \boldsymbol{0} \right), \label{Collinear frame q}
\end{align}
where  \( P^2 = M^2 \), and \( \kappa \) is the Nachtmann variable \cite{Nachtmann:1973mr} defined through the following relation
\begin{equation}
    2 P \cdot q = \frac{Q^2}{\kappa} - \kappa M^2.
\end{equation}
In the approximation where nucleon mass corrections are neglected, \( \kappa \) reduces to the Bjorken-\( x \) variable. Using \( k_1^+ = \xi P^+ \), we can express the four-vector \( k_1^\mu \) as
\begin{equation}
    k_1^\mu = \left( \xi P^+, \frac{m_1^2}{2\xi P^+}, \boldsymbol{0} \right).
\end{equation}
With this, and utilizing the relations for \( P \) and \( q\) in \cref{Collinear frame P,Collinear frame q}, together with the metric defined earlier in  \cref{Collinear frame metric}, we obtain the scalar products that are necessary for deriving the results in \cref{eqn: H4 mixing,eqn:H5 mixing} 

\begin{align}
   & k_1 \cdot q = \frac{\xi Q^2}{2x} - \frac{m_1^2 x}{2\xi}, \nonumber\\
   &  q \cdot P = \frac{Q^2}{2x}, \nonumber\\
   & k_1 \cdot P = \frac{m_1^2}{2\xi}.
   \label{scalar products}
\end{align}

Keeping the parton mass \(m_1\) explicit is crucial, as it underlies the mixing of partonic structure functions and gives rise to the corresponding terms in \cref{eqn: H4 mixing,eqn:H5 mixing}.

\section{\texorpdfstring{$\bm{\hat{\mathcal{F}}_{2}^{\mathrm{QS, \, NLO}}}$}{F2 QS}}

The partonic structure function $\hat{F}_{2}^{\rm QS}$ appears in the mixing terms within the expressions provided in \cref{eqn: H4 mixing,eqn:H5 mixing}. The analytical expression for this function is given by \cite{Kretzer:1998ju}:
\begin{equation}
\begin{aligned}
\hat{F}_{2}^{\rm QS}(\hat{s}_1,\hat{t}_1)  & = 8 \frac{ Q^2}{ \hat{x} \Delta^{\prime 4}}\left\{-2 \Delta^4 q_{+}\left(\frac{m_2^2}{\hat{s}_1^2}+\frac{m_1^2}{\hat{t}_1^2}+\frac{\Sigma_{++}}{\hat{s}_1 \hat{t}_1}\right)+2 m_1 m_2 q_{-}\left(\frac{\left(\Delta^{\prime 2}-6 m_1^2 Q^2\right) \hat{s}_1}{\hat{t}_1}\right.\right. \\
& \left.+2\left(\Delta^{\prime 2}-3 Q^2\left(\hat{s}_1+\Sigma_{++}\right)\right)+\hat{t}_1 \frac{\Delta^{\prime 2}-6 Q^2\left(m_2^2+\hat{s}_1\right)}{\hat{s}_1}\right)\\
& +q_{+}\left(\frac{-2 m_1^2 \hat{s}_1\left[\left(\Delta^2-6 m_1^2 Q^2\right) \hat{s}_1+2 \Delta^2 \Sigma_{+-}\right]}{\hat{t}_1^2}+\frac{-2 \Delta^2\left(\Delta^2+2 \Sigma_{+-} \Sigma_{++}\right)}{\hat{t}_1}\right. \\
& +\frac{-\hat{s}_1\left[2\left(\Delta^2-6 m_1^2 Q^2\right) \hat{s}_1+\left(\Delta^{\prime 2}-18 m_1^2 Q^2\right) \Sigma_{++}+2 \Delta^2\left(3 \Sigma_{++}-4 m_1^2\right)\right]}{\hat{t}_1} \\
& +\left[-2\left(m_1^2+m_2^2\right) \hat{s}_1^2-9 m_2^2 \Sigma_{+-}^2-\frac{2 \Delta^2\left(\Delta^2+2 m_2^2 \Sigma_{+-}\right)}{\hat{s}_1}+2 \hat{s}_1\left[2 \Delta^2\right.\right. \\
& \left.\left.\left.\left.+\left(m_1^2-5 m_2^2\right) \Sigma_{+-}\right]+\Delta^2\left(2 \Sigma_{++}-m_2^2\right)\right]-\hat{t}_1 \frac{\left[\Delta^{\prime 2}-6 Q^2\left(m_2^2+\hat{s}_1\right)\right] \Sigma_{++}}{\hat{s}_1}\right)\right\},
\end{aligned}
\end{equation}
from which we obtain
\begin{equation}
\hat{\mathcal{F}}_2^{\mathrm{QS, \, NLO}}(\xi^\prime)  =  C_F \left( N_2 (S_2 + V_2) \delta(1 - \xi') + \frac{1}{8} \frac{1 - \xi'}{(1 - \xi')_+} \frac{\hat{s}_1}{\hat{s}_1 + m_2^2} \hat{F}_{2}^{\rm QS}(\hat{s}_1)\right).
\end{equation}
The coefficients are given by
\begin{equation}
   V_2= \left(C_{R,-}+ \frac{1}{2}(m_{2}^2 C_{1,-}+ m_{1}^2 C_{1,+})\right)+ \frac{q_-}{ q_+} \left( \frac{1}{2} m_1 m_2 (C_{1,-}+C_{1,+})+ C_+\right)  ,\quad  S_2= S_5, \quad  N_2=N_5.
\end{equation}

\section{Plus-distribution identities}
\label{appendix: Plus distribution identities}
This appendix collects useful properties of plus distributions, which are essential for manipulating convolution kernels in the main text. We use the following definition \cite{Field:1989uq}:
\begin{equation}
  [p(x)]_+ = \lim_{\epsilon\to 0}\left(\theta(1-\epsilon-x)\,p(x)
  - \delta(1-\epsilon-x)\int_0^{1-\epsilon}dy\,p(y)\right).
  \label{eqn: plusdef}
\end{equation}
and the plus distribution over \([\chi, 1]\) with weight \(s\) is defined as:
\begin{align}
 \Big([p]_+ s\,\otimes f\Big)(\chi) 
 := \int_{\chi}^{1} dz\, [p(z)]_+ s(z)
 f\!\left(\tfrac{\chi}{z}\right).
\end{align}
From the definition, the following identities hold:
\begin{align}
(A(x)+B(x))|_+ &= A(x)|_+ + B(x)|_+, \\
(c\,p(x))|_+ &= c\,p(x)|_+ \quad (c\in\mathbb{R}), \\
(p(x)|_+)|_+ &= p(x)|_+.
\end{align}
These reflect linearity, homogeneity, and idempotence.

The following lemmas are key for our work:

\begin{lemma}[Convolution Identity]\label{lem:convolution}
The convolution of a plus distribution with weight \(s(z)\) over \([\chi, 1]\) satisfies:
\begin{align}
 \big([p]_+ s \otimes f\big)(\chi) 
 = \int_{\chi}^{1} dz\,p(z)\Big[s(z) f\!\left(\frac{\chi}{z}\right) - s(1)f(\chi)\Big]
 - s(1)f(\chi)\int_0^{\chi}dz\,p(z).
 \label{eqn: :main-identity}
\end{align}
\end{lemma}

\begin{proof}
\begin{align}
 \left[\left([p]_+ s\right)\otimes f\right](\chi) \overset{\text{def}}{=} &
    \int_{\chi}^{1} d z \, [p(z)]_+ s(z) f\left( \frac{\chi}{z} \right) \nonumber \\
    = & \int_{\chi}^{1} dz \, \lim_{\epsilon \rightarrow 0} \left( \theta(1-\epsilon-z) p(z) - \delta(1-\epsilon-z) \int_{0}^{1-\epsilon} dx \, p(x) \right) s(z) f\left( \frac{\chi}{z} \right)  \nonumber \\ 
   = & \lim_{\epsilon \rightarrow 0} \int_{\tilde{\chi}}^{1-\epsilon} dz \, p(z) s(z) f\left( \frac{\chi}{z} \right)  - \int_{\chi}^{1} dz \, \lim_{\epsilon \rightarrow 0} \delta(1-\epsilon-z) \int_{0}^{1-\epsilon} dx \, p(x)  s(z) f\left( \frac{\chi}{z} \right) \nonumber\\
   = &  \int_{\chi}^{1} dz \, p(z) \left(s(z) f\left( \frac{\chi}{z} \right)- s(1) f( \chi)\right) + s(1) f(\chi) \lim_{\epsilon \rightarrow 0} \int_{\chi}^{1-\epsilon} dz \,  p(z)  \nonumber\\
   & - \lim_{\epsilon \rightarrow 0} \int_{0}^{1-\epsilon} dx  \, p(x) \int_{\tilde{\chi}}^{1} dz \, \delta(1-\epsilon-z)  s(z) f\left( \frac{\chi}{z} \right)  \nonumber\\
    = & \int_{\chi}^{1} dz \, p(z) \left(s(z) f\left( \frac{\chi}{z} \right)- s(1) f( \chi)\right) + s(1) f(\chi) \lim_{\epsilon \rightarrow 0} \int_{\chi}^{1-\epsilon}  dz \, p(z)  \nonumber\\
   & - \lim_{\epsilon \rightarrow 0} \int_{0}^{1-\epsilon} dx  \, p(x) s(1-\epsilon) f \left( \frac{\chi}{1-\epsilon} \right) \nonumber \\
    =&  \int_{\chi}^{1} dz \,  p(z) \left(s(z) f\left( \frac{\chi}{z} \right)- s(1) f( \chi)\right) + s(1) f(\chi) \lim_{\epsilon \rightarrow 0} \int_{\chi}^{1-\epsilon} dz \,  p(z)  \nonumber\\
   & -\left( s(1) f(\chi) \lim_{\epsilon \rightarrow 0} \int_{\chi}^{1-\epsilon} dz  \, p(z)  + s(1) f(\chi) \int_{0}^{\chi} dz \, p(z)  \right) \nonumber
   \\
    = & \int_{\chi}^{1} dz\,  p(z) \left(s(z) f\left( \frac{\chi}{z} \right)- s(1) f( \chi) \right)  -s(1) f(\chi) \int_{0}^{\chi} dz \,  p(z) .
\end{align}
\end{proof}

\begin{lemma}[Product Rule]
Let \(s_1(x)\) be smooth at \(x=1\), and let \(p_1(x)\) be a kernel with plus prescription. Then, as distributions:
\begin{equation}
  (p_1(x)s_1(x))\big|_+ 
  \,=\, s_1(x)\,p_1(x)\big|_+ 
  \, +\, \delta(1-x)\int_0^1 dy\,p_1(y)\big[s_1(1)-s_1(y)\big].
  \label{eqn:lemma-product}
\end{equation}
\end{lemma}

\begin{proof}
Define \(P(x):=p_1(x)s_1(x)\). The convolution is
\begin{align*}
(P\otimes f)(x)&=\int_x^1dy\,p_1(y)s_1(y)[f(x/y)-f(x)]-f(x)\int_0^xdy\,p_1(y)s_1(y).
\end{align*}
The weighted plus convolution is
\begin{align*}
(p_1|_+\,s_1)\otimes f(x)&=\int_x^1dy\,p_1(y)[s_1(y)f(x/y)-s_1(1)f(x)]-f(x)s_1(1)\int_0^xdy\,p_1(y).
\end{align*}
Subtracting gives
\begin{align*}
&(P\otimes f)(x)-(p_1|_+\,s_1)\otimes f(x)\\
&= f(x)\Bigg[\int_x^1dy\,p_1(y)(s_1(1)-s_1(y)) + \int_0^xdy\,p_1(y)(s_1(1)-s_1(y))\Bigg]\\
&= f(x)\int_0^1dy\,p_1(y)(s_1(1)-s_1(y)),
\end{align*}
which is a \(\delta(1-x)\) term with the coefficient in \cref{eqn:lemma-product}.
\end{proof}

\section{Numerical implementation of the plus distribution}
\label{appendix: Plus Distribution Treatment}

This appendix details the numerical treatment of the convolution integrals appearing in the NLO QS processes. The accurate evaluation of these integrals requires a careful handling of the plus distribution arising in the hard scattering Wilson coefficients.

The convolution integrals for the structure functions are expressed as
\begin{align}
     F_{i,\, \alpha, \beta }^{\mathrm{QS}, \, \mathrm{NLO}, \, k}(x,Q^2)
    &= 
     \frac{\alpha_s}{2 \pi} 
       \int_\chi^1 d \xi^{\prime} \,
       f_\alpha\!\left(\frac{\chi}{\xi^{\prime}}, Q^2\right)
       \mathcal{H}_{i, \alpha, \beta}^{\mathrm{QS}, \mathrm{NLO}, \, k}(\xi^\prime),
    \label{eqn: qs convolution} \\
    \mathcal{H}_{i, \alpha, \beta}^{\rm QS, \, NLO, \, k}(\xi^\prime)
    &= L_{i , \alpha, \beta}^k \, \delta(1-\xi^\prime)
       + \left[p_{1}(\xi^\prime)\right]_+ S_{1,\, i, \alpha , \beta}^{k} (\xi^\prime),
    \label{eqn: QS convolutions}
\end{align}
where $p_1(\xi^\prime)=1/(1-\xi^\prime)$ and
\begin{align}
S_{1,4, \alpha, \beta}^k(\xi^\prime) &=   \mathcal{K} \left( \hat{F}_{4, \alpha, \beta}^{\text{QS},k}(\xi^\prime)   + (\mathcal{R}_2-1) \left( \mathcal{R}_1\hat{F}_{2, \alpha, \beta}^{\text{QS},k}(\xi^\prime)  +\hat{F}_{5, \alpha, \beta}^{\text{QS},k}(\xi^\prime) \right)\right)
\\
S_{1,5, \alpha, \beta}^k(\xi^\prime) &= \mathcal{K}  \mathcal{R}_2\left(\hat{F}_{5, \alpha, \beta}^{\text{QS},k}(\xi^\prime)+ 2 \mathcal{R}_1 \hat{F}_{2, \alpha, \beta}^{\text{QS}, \, k}(\xi^\prime)\right)
\label{mixing equations}
\end{align}
with
\begin{equation}
    \mathcal{K}= \frac{1}{8} (1-\xi^\prime)\frac{\hat{s}_1}{\hat{s}_1+m_2^2}.
\end{equation}
The coefficients $L_{i, \alpha, \beta}^k$ are given by
\begin{align}
L_{2, \alpha, \beta}^k &= \chi \frac{(1+\bar{\mathcal{R}}_3)^2}{(1-\bar{\mathcal{R}}_3)} \hat{N}_{2,\alpha, \beta}^k, \\
L_{4, \alpha, \beta}^k &= \hat{N}_{4,\alpha, \beta}^k + (\bar{\mathcal{R}}_2 - 1)\big( \bar{\mathcal{R}}_1 \hat{N}_{2,\alpha, \beta}^k + \hat{N}_{5,\alpha, \beta}^k \big), \\
L_{5, \alpha, \beta}^k &= \bar{\mathcal{R}}_2\big( \hat{N}_{5, \alpha, \beta}^k + 2 \bar{\mathcal{R}}_1 \hat{N}_{2, \alpha,\beta}^k \big),
\label{mixing equations}
\end{align}
where $\bar{R}_i$ are the $R_i$ coefficients evaluated at $\xi^\prime=1$, and
\begin{align}
\hat{N}_{2, \alpha, \beta}^k &= N_{2, \alpha, \beta}^k ( V_{2, \alpha, \beta} + S_{2, \alpha, \beta}), \\
\hat{N}_{4, \alpha, \beta}^k &= N_{4, \alpha, \beta}^k V_{4,\alpha, \beta}, \\
\hat{N}_{5,\alpha, \beta}^k &= N_{5,\alpha,\beta}^k ( V_{5, \alpha, \beta} + S_{5, \alpha, \beta}).
\end{align}
Note that $S_4 = 0$ and that we use the index convention of \cref{eqn:index_convention_coefficients}.

Using the identities for the plus distribution, the convolution products can be expressed as follows, which also directly corresponds to their implementation in the code:
\begin{equation}
     F_{i,\alpha, \beta}^{\text{QS, NLO}, \, k}(x,Q^2)
    = 
    \frac{\alpha_s}{2 \pi} 
      \left[
        \mathcal{L}_{1,i, \alpha, \beta}^k f_\alpha(\chi)
        + \int_{\chi}^{1} \frac{S_{1,i, \alpha, \beta}^k( \xi^\prime)}{1-\xi^\prime}
          \left( f_\alpha \left( \frac{\chi}{\xi^\prime}\right) - f_\alpha(\chi) \right)
        d\xi^\prime
      \right],
\end{equation}
with
\begin{align}
\mathcal{L}_{1,i, \alpha, \beta}^k
= L_{i, \alpha, \beta}^k
+ \ln(1-\chi) \bar{S}_{1, \, i, \alpha, \beta}^k
+ \int_{\chi}^{1} \frac{S_{1, i, \alpha, \beta}^k(\xi^\prime)-\bar{S}_{1,i , \alpha, \beta}^k}{1-\xi^\prime} \, d \xi^\prime.
\label{eqn:sudakov log}
\end{align}
The coefficients $\bar{S}_{1,\alpha, \beta}^k$ are given by
\begin{align}
\bar{S}_{1,1, \alpha, \beta}^k &= \chi \frac{(1+\bar{\mathcal{R}}_3)^2}{(1-\bar{\mathcal{R}}_3)} \hat{S}_{1,1, \alpha, \beta}^k, \\
\bar{S}_{1,4, \alpha, \beta}^k &= \hat{S}_{1,4, \alpha, \beta}^k + (\bar{R}_2 - 1)(\bar{R}_1 \hat{S}_{1,2, \alpha, \beta}^k + \hat{S}_{1,5, \alpha, \beta}^k), \\
\bar{S}_{1,5, \alpha, \beta}^k &= \bar{\mathcal{R}}_2(\hat{S}_{1,5, \alpha, \beta}^k + 2 \bar{\mathcal{R}}_1 \hat{S}_{1,2, \alpha, \beta}^k).
\end{align}
where
\begin{align}
  \hat{S}_{1,2, \alpha, \beta}^k &= -2 N_{2, \alpha, \beta}^k (2 - \Sigma_{++} I_1), \\
  \hat{S}_{1,4, \alpha, \beta}^k &= 0, \\
  \hat{S}_{1,5, \alpha, \beta}^k &= -2 N_{5, \alpha, \beta}^k (2 - \Sigma_{++} I_1),
\end{align}
Note that the coefficients $\hat{S}_{1,i}$ can also be derived from

\begin{align}
\hat{S}_{1, i, \alpha, \beta}^k
&= \lim_{\hat{s}_1 \rightarrow 0} \left[
   -2 N_{i, \alpha, \beta}^k
   \frac{\hat{s}_1^2}{m_{2}^2}
   \int_{0}^{1} \left(
      \frac{m_{2}^2}{\hat{s}_{1}^2}
      + \frac{m_1^2}{\hat{t}_1^2}
      + \frac{\Sigma_{++}}{\hat{s}_1 \hat{t}_1}
   \right) dy
   \right] \\
&= -2 N_{i, \alpha, \beta}^k (2 - \Sigma_{++} I_1).
\end{align}

Following the same logic for the subtraction term:
\begin{align}
 F_{i, \alpha, \beta}^{\mathrm{QS,\, SUB},\,k}(x,Q^2)
&= \mathcal{N}_{i, \alpha, \beta}^{\mathrm{QS,\, LO},\, k}
   \times  \frac{\alpha_s}{2\pi} 
   \int_{\chi}^1 d\xi'\;
   f_\alpha\!\left(\frac{\chi}{\xi'}, \mu^2\right)
   \mathcal{H}_{i,\alpha}^{\text{QS, SUB}, k}(\xi^\prime), \\
\mathcal{H}_{i,\alpha}^{\text{QS, SUB}, k} (\xi^\prime)
&= c_i
   C_F \left[
     \frac{1+\xi'^2}{1-\xi'}
     \left(\ln \frac{Q^2}{m_\alpha^2} - 1 - 2\ln(1-\xi')\right)
   \right]_+.
\label{eqn: QS sub}
\end{align}
with $c_4=0, \, c_5=1$ we can write:
\begin{equation}
 F_{i, \alpha, \beta}^{\mathrm{QS, SUB}, k}(x,Q^2)
= \mathcal{N}_{i, \alpha, \beta}^{\mathrm{QS,\, LO},\, k}
  \times  \frac{\alpha_s}{2 \pi} 
  \left[
    P_{2,i,  \alpha}(\chi)f(\chi)
    + \int_{\chi}^{1} p_{2, i, \alpha}(\xi^\prime)
      \left( f_\alpha \left( \frac{\chi}{\xi^\prime} \right) - f_\alpha(\chi) \right)
    d\xi^\prime
  \right],
\end{equation}
where
\begin{align}
p_{2, i, \alpha}(\xi^\prime)
&= c_i C_F
  \frac{1+\xi'^2}{1-\xi'}
  \left(\ln \frac{Q^2}{m_\alpha^2} - 1 - 2\ln(1-\xi')\right), \\
P_{2,i,  \alpha}(\chi)
&= - \int_{0}^{\chi} p_{2, i, \alpha}(\xi^\prime) \, d\xi^\prime. \nonumber\\
&= c_i C_F \left[2\chi 
+ \ln\!\left(\frac{Q^{2}}{m_\alpha^{2}}\right)
 \left( \frac{\chi(2+\chi)}{2} + 2\ln(1-\chi) \right)
- \ln(1-\chi)\left( -1 + \chi(2+\chi) + 2\ln(1-\chi) \right)\right]
\label{eqn:sudakov close to 1 sub}
\end{align}
Note the emergence of the logarithmic terms $\ln(1-\chi)$ and $\ln^2(1-\chi)$ in \cref{eqn:sudakov log,eqn:sudakov close to 1 sub} corresponds to the soft-gluon instabilities previously outlined in the main text, \textit{cf.} \cref{sec:numerical}.


\bibliography{polACOT_references}

\end{document}